\documentclass[journal]{IEEEtran}

\usepackage{graphicx}
\usepackage[dvipsnames]{xcolor} 
\usepackage{tabularx}
\usepackage[compress,nospace]{cite}
\usepackage{bchart}
\usepackage{booktabs}
\usepackage{multirow}
\usepackage{amssymb}
\usepackage{pifont}
\usepackage{array,colortbl}
\usepackage{adjustbox}
\usepackage{amsfonts}
\usepackage{multirow}
\usepackage{tikz}
\usepackage{float}
\usepackage{textcomp}
\usepackage{braket}
\usepackage{amsmath,calc}
\usepackage{subcaption}
\usepackage{hhline}
\usepackage{array, makecell}
\usepackage{boldline}
\usepackage{caption}
\usepackage{booktabs}
\usepackage[flushleft]{threeparttable}
\usepackage{amsmath}
\usepackage[utf8]{inputenc}
\usepackage{tikz}
\usetikzlibrary{shapes.geometric, arrows}
\usepackage{colortbl}
\usepackage{fixltx2e}
\usepackage{subcaption}
\usepackage{enumitem}
\usepackage{textcomp}

\graphicspath{{./figures/}}

\DeclareCaptionLabelFormat{andtable}{#1˜#2 \& \tablename˜\thetable}

\newcolumntype{R}[2]{%
	>{\adjustbox{angle=#1,lap=\width-(#2)}\bgroup}%
	l%
	<{\egroup}%
}

\makeatletter
\newcommand*{\rom}[1]{\expandafter\@slowromancap\romannumeral #1@}
\makeatother

\makeatletter
\def\endthebibliography{%
	\def\@noitemerr{\@latex@warning{Empty `thebibliography' environment}}%
	\endlist
}
\makeatother

\begin{document}
	
	\title{Sequential Pattern mining of  Longitudinal  \\ 
		Adverse Events After Left Ventricular \\  
		Assist  Device Implant}

	\author{ Faezeh Movahedi\thanks{Faezeh Movahedi is with Department of Electrical and Computer Engineering, Swanson School of Enginering, University of Pittsburgh, Pittsburgh, PA, USA. E-mail: fam32@pitt.edu.},  Robert L. Kormos\thanks{ Robert L. Kormos and Laura Seese are with Heart and Vascular Institute, University of Pittsburgh Medical Center, Pittsburgh, PA, USA. Emails: seeselm@upmc.edu, kormosrl@upmc.edu.}, Lisa Lohmueller\thanks{Lisa Lohmueller is with School of Computer Science, Carnegie Mellon University,Pittsburgh, PA, USA. E-mail: lcarey@cmu.edu.}, Laura Seese, Manreet Kanwar\thanks{Manreet Kanwar and  Srinivas Murali are with the Cardiovascular Institute at Allegheny General Hospital, Pittsburgh, PA, USA. Emails: Manreet.KANWAR@ahn.org, Srinivas.MURALI@ahn.org.},  \\
		Srinivas Murali,  Yiye Zhang\thanks{Yiye Zhang is  with Weill Cornell medical school, University of Cornell University, New York, NY, USA. Email: yiz2014@med.cornell.edu.}, Rema Padman\thanks{Rema Padman is with Heinz college, Carnegie Mellon University, Pittsburgh, PA, USA. Email:rpadman@cmu.edu.},  James F.  Antaki\thanks{James F Antaki is  with Biomedical Engineering Department, University of Cornell University, Ithaca, NY, USA. Email: antaki@cornell.edu.}}
	
	\date{}
	\maketitle
	
	\begin{abstract}
		Left ventricular assist devices (LVADs) are an increasingly common therapy for patients with advanced heart failure. However, implantation of the LVAD increases the risk of stroke, infection, bleeding, and other serious adverse events (AEs). Most post-LVAD AEs studies have focused on individual AEs in isolation, neglecting the possible interrelation, or causality between AEs. This study is the first to conduct an exploratory analysis to discover common sequential chains of AEs following LVAD implantation that are correlated with important clinical outcomes. This analysis was derived from 58,575 recorded AEs for 13,192 patients in International Registry for Mechanical Circulatory Support (INTERMACS) who received a continuous-flow LVAD between 2006 and 2015. The pattern mining procedure involved three main steps: (1) creating a bank of AE sequences by converting the AEs for each patient into a single, chronologically sequenced record, (2) grouping patients with similar AE sequences using hierarchical clustering, and (3) extracting temporal chains of AEs for each group of patients using Markov modeling. The mined results indicate the existence of seven groups of sequential chains of AEs, characterized by common types of AEs that occurred in a unique order. The groups were identified as: GRP1: Recurrent bleeding, GRP2: Trajectory of device malfunction \& explant, GRP3: Infection, GRP4: Trajectories to transplant, GRP5: Cardiac arrhythmia, GRP6: Trajectory of neurological dysfunction \& death, and GRP7: Trajectory of respiratory failure, renal dysfunction \&	death. These patterns of sequential post-LVAD AEs disclose potential interdependence between AEs and may aid prediction, and prevention, of subsequent AEs in future studies. 
		
	\end{abstract}

	\section{Introduction}
	
	\begin{figure}[ht]
		\centering
		\includegraphics[width=0.2\textwidth,height=0.2\textheight]{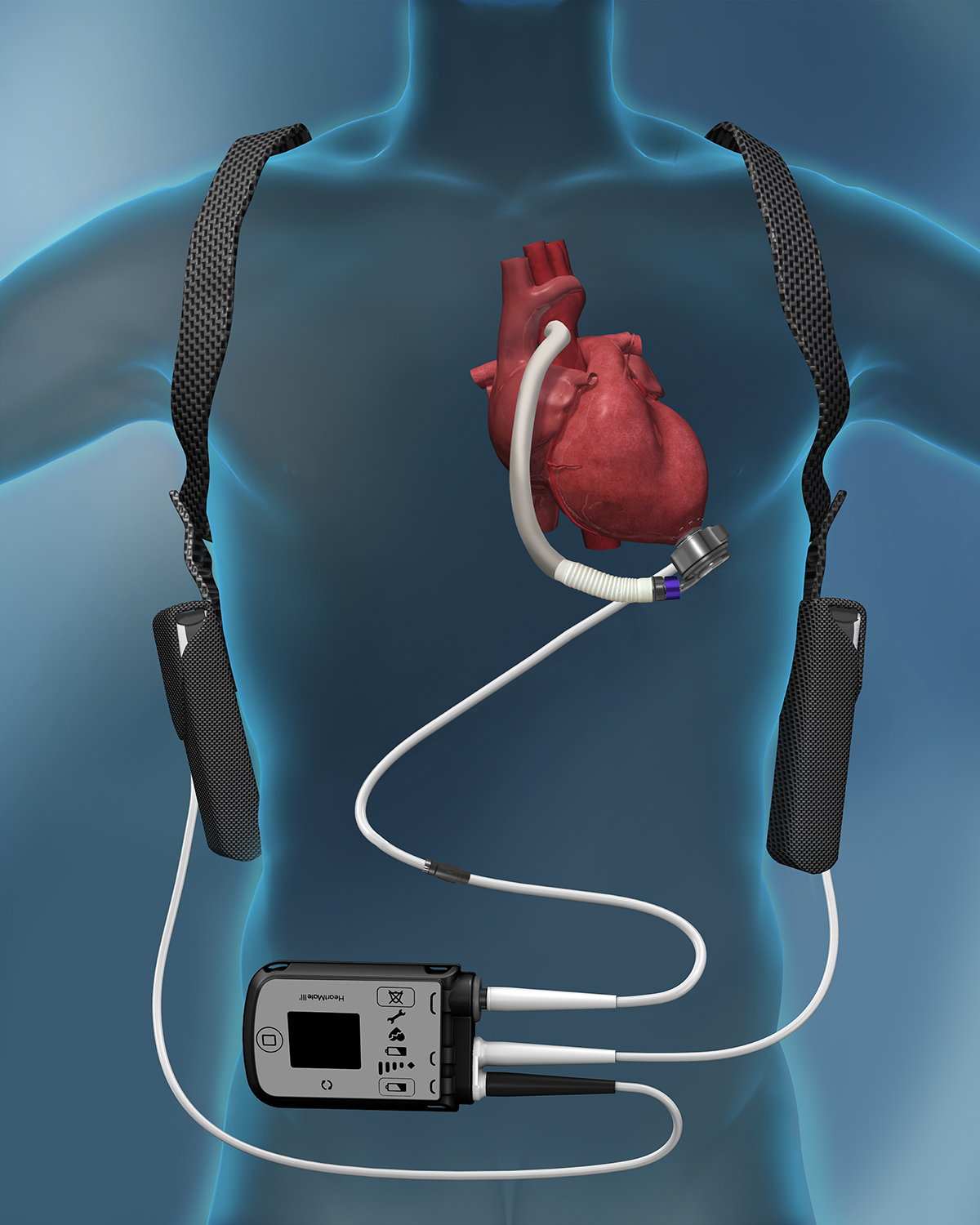}
		\caption{Illustration of HeartMate 3 LVAD from Thoratec Corporation \cite{LVADThoratec}. LVAD is a blood pump that is connected to external components out of the body, including a controller and power sources worn by patients, by a driveline.}
	\end{figure}
	
	\IEEEPARstart{P}{atients} with advanced congestive heart failure (CHF) do not respond to traditional therapies like medical treatment and symptom management. Heart transplantation is considered the gold standard treatment for eligible patients with advanced CHF but is severely limited by donor availability. It was performed for only 2000-2500 eligible patients per year in the US (half of the number of candidates added to the wait list every year)  \cite{khush2015national}.  Therefore, there has been an increasing need to provide advanced CHF patients with alternative treatment options,such as left ventricular assist devices (LVAD). The LVAD is a pump that augments the flow of blood from the left ventricle of the heart, as shown in Fig. 1.
	
	
	In 2005, the National Institutes of Health (NIH) established INTERMACS, which is currently the nation-wide database of longitudinal data for mechanical assist devices with data from $>$20,000 patients and $>$180 hospitals over the past decade \cite{kirklin2017eighth}. Based on the INTERMACS 2017 annual report, overall 1-year and 2- year patient survival rates were 81$\%$ and 70$\%$, respectively \cite{kirklin2017eighth}. It also reported an enhanced patient quality of life in the first 3 months after LVAD implantation that lasted for at least 24 months \cite{kirklin2017eighth}. Yet, INTERMACS reported several risks of recurrent adverse events (AEs) following LVAD implantation including bleeding, infection, cardiac arrhythmia, stroke, and respiratory failure \cite{kirklin2017eighth}.  For 17,632 LVAD patients between 2008- 2016, bleeding and infection were the most frequent AEs, especially in the first 3 months post-LVAD with rate (events per 100 patient-months) of 16.24 and 13.63, respectively. More importantly, subsequent survival is impacted by the incidence of major AEs during the first 3 months post-LVAD \cite{kirklin2017eighth}. 
	
	\begin{figure*}[]
		\centering
		\includegraphics[width=0.8\textwidth,height=0.18\textheight]{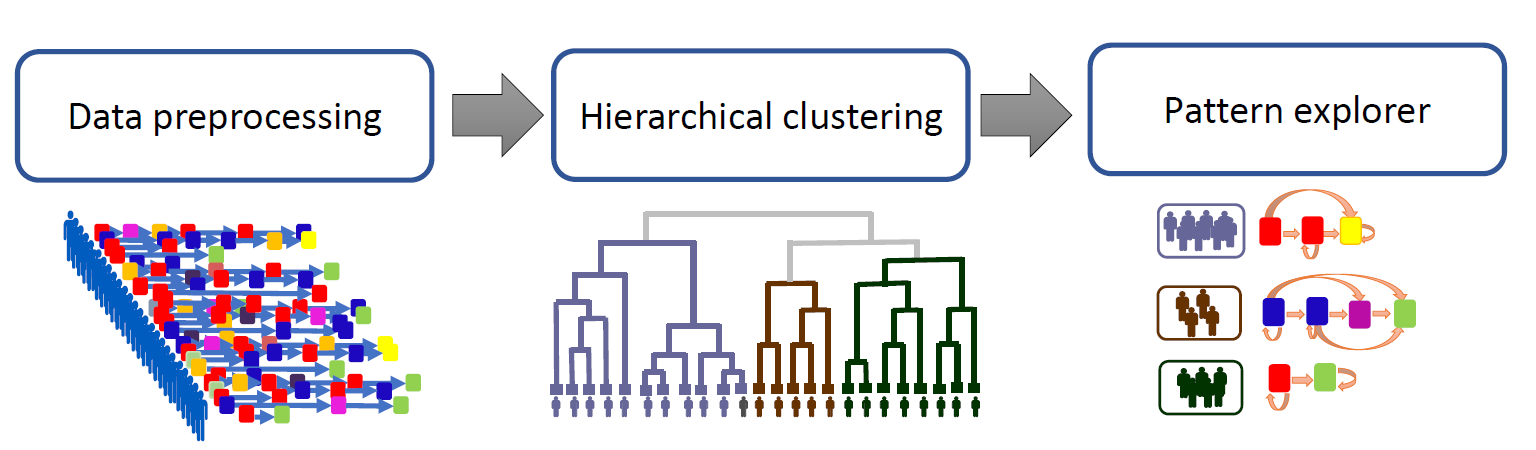}
		\caption{General work-flow: 1. Data preprocessing:Forming patients' sequences. 2. Hierarchical clustering:cluster patients into groups of patients with high similarity between patients' sequences. 3. Pattern explorer: Extracting patterns of post-LVAD sequential AEs in each group using Markov Molding \cite{zhang2015paving}. }
	\end{figure*}

	There have been numerous clinical studies in recent years focusing on the incidence, prevalence, and risk factors of post-LVAD AE's \cite{slaughter2009advanced,joy2016risk,draper2014gi,stulak2016adverse,bunte2013major,topkara2010infectious,harvey2015stroke,willey2016outcomes,sami2014gastrointestinal,boyle2014pre,toda2012risk,goldstein2012continuous}. Most studies emphasized one type of major AE like bleeding \cite{joy2016risk,bunte2013major,sami2014gastrointestinal,draper2014gi}, infection \cite{toda2012risk,goldstein2012continuous,schaffer2011infectious,sharma2012driveline,topkara2010infectious}, and stroke \cite{slaughter2009advanced,boyle2014pre,harvey2015stroke,willey2016outcomes,al2016neuroendovascular,kato2012pre}. Some studies investigated AEs of one specific type of LVAD pump \cite{stulak2016adverse,bunte2013major,boyle2014pre,goldstein2012continuous}, while others compared AEs between different types of pumps  \cite{slaughter2009advanced,joy2016risk,schaffer2011infectious,draper2014gi}. Some studies underlined the difference between the AEs of LVAD as short-term versus long-term therapy \cite{boyle2014pre,goldstein2012continuous}. There have been also few studies investigated data-driven  techniques for AEs  prediction including applying Bayesian model to predict the right heart failure or survival after LVAD \cite{kanwar2019predicting,kanwar2018bayesian} and developing decision support system for patients management after LVAD \cite{karvounis2014decision}. 
	

	Although these studies provide many valuable clinical aspects of AEs, they suffer from major limitations. Some of the results from these studies cannot be generalized as they were case studies or based on one hospital with low numbers of patients, usually a few hundred patient, compared  with thousands of patients in INTERMACS. Furthermore, most were clinical studies that focused on traditional  statistical methods such as multivariable Cox-Regression Models to establish hazard ratios, odds ratios, and event rates. No study has implemented  modern data mining techniques to model \textit{patterns} of post-LVAD AEs. Some are cross sectional studies based on data at a specific point in time after LVAD implantation and therefore do not consider the transitions between AEs over  duration after LVAD implant. In addition, these studies treated each type of AE as a separate event, neglecting the possible interrelation or bootstrapping between AEs. LVAD-associated AE rarely occur in isolation. This is due to the causes of events being closely related. For example, a single gastrointestinal (GI) bleed is often followed by more GI bleeds, as the introduction of the continuous flow LVAD can lead to von Willebrand factor dysfunction and arteriovenous malformations \cite{sami2014gastrointestinal}. Similarly, bleeding and stroke can occur one after the other in LVAD patients due to issues finding the right balance of anticoagulation \cite{shah2017bleeding}.  No research has studied the pattern of transition between AEs to answer questions such as, “Which AEs  occur commonly in a specific temporal order after LVAD implantation?”.
	
	
	A patient’s post-LVAD AEs can be represented as a temporal sequence in which each AE is considered a unique element connected to other elements in a sequence to identify the transitions between AEs. Data from a large number of patients’ histories of temporal AEs could be summarized and visualized as a set of temporal sequences of AEs and then patterns within the AE sequences could be extracted using data mining techniques. Several recent studies have applied various temporal mining techniques, such as Markov models and Heuristics models,  to extract recurring patterns from longitudinal healthcare data \cite{rojas2016process,yang2014process,ghasemi2016process,mans2008application,kaymak2012process,zhou2014process,augusto2016evaluation,fernandez2014temporal,erdogan2018systematic}. These studies have identified possible relativity and association between clinical variables such as medical interventions, diagnoses, and treatments \cite{rojas2016process,yang2014process,ghasemi2016process,mans2008application,kaymak2012process,zhou2014process,augusto2016evaluation,fernandez2014temporal,erdogan2018systematic}.
	
	To the best of our knowledge, this study is the first attempt to explore and model sequential patterns of post-LVAD AEs  and specifically differentiate distinct groups of patients based on their sequences of AEs. The overall approach for this study is motivated by the methodology introduced by Zhang, et al.  \textit{et al.} \cite{zhang2015paving,zhang2015clinical} and has three main steps, as shown in Fig. 2. First, selected data from INTERMACS were transferred into sequences of AEs for each patient through multiple preprocessing tasks. Next, patients' AE sequences were clustered into groups with similar sequences using hierarchical clustering.  Lastly, patterns of chains of transitions between AEs for each group were extracted using Markov Modeling. The results of this study provide valuable real-world insights about post-LVAD AEs patterns, which clinicians can use to make more informed treatment and management decisions to improve clinical outcomes of LVADs.

	\section{Methodology}
	
	\subsection{Data pre-processing}
	
	\subsubsection{INTERMACS data selection and pre-processing}  
	
	This study included 58,575 recorded AEs of 13,192 patients (median age of 50-59; 10,333 male vs. 2,859 female;) with advanced heart failure who received a continuous flow LVAD between 2006 to 2015, extracted from INTERMACS. It should be noted that the incidence of heart failure is significantly higher in men than women at all ages \cite{mehta2006gender}. For patients with multiple device implants, AEs after the first LVAD explant are excluded as patients with multiple subsequent LVAD devices are clinically treated differently.
	
	\begin{figure}[ht]
		\centering
		\includegraphics[width=0.4\textwidth,height=0.18\textheight]{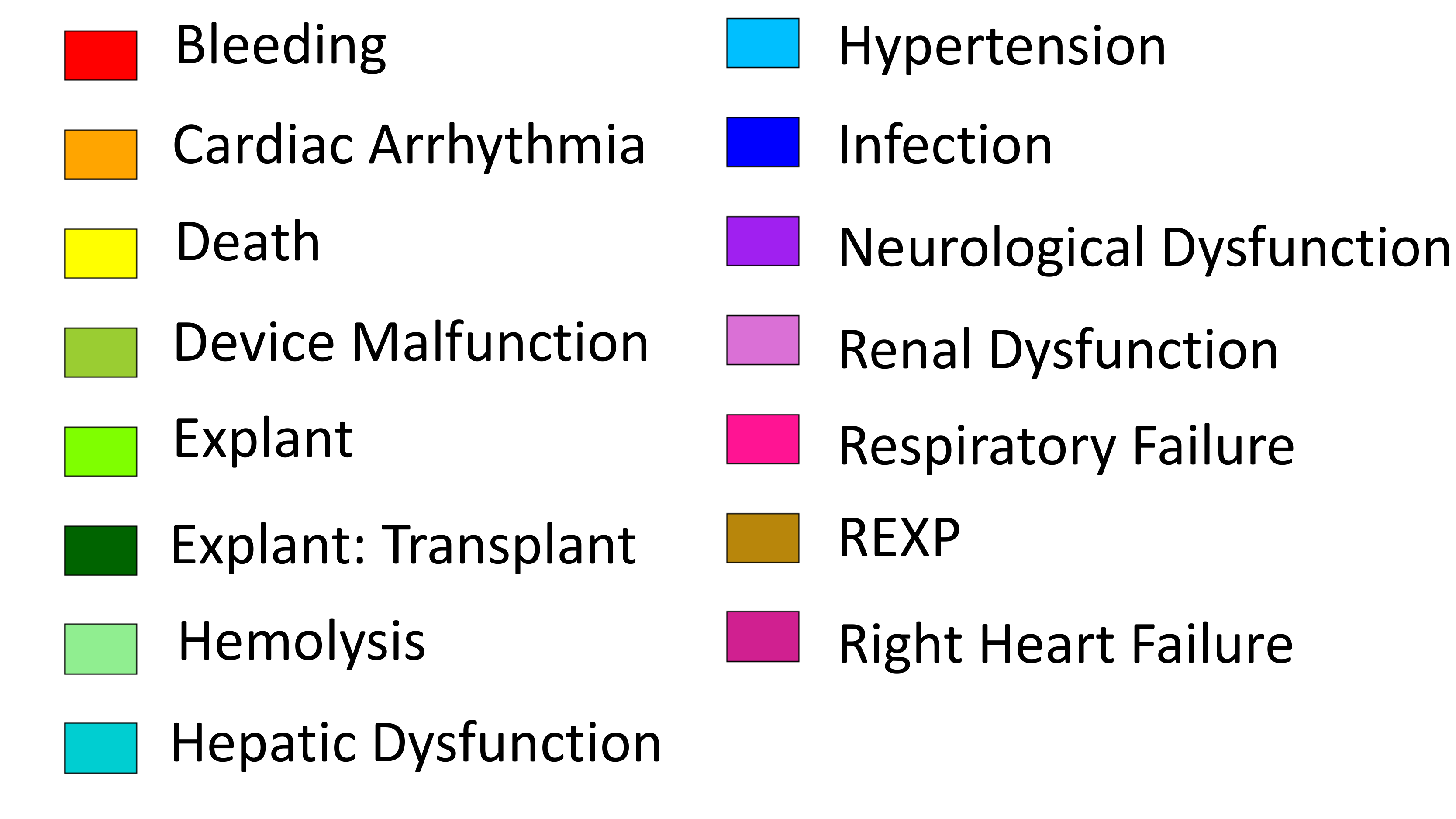}
		\caption{Color code for 15 types of AEs and final outcomes}
	\end{figure}

	Final outcomes, such as death, explant, and transplant, were included as the last elements in the sequences of AEs. For the subset of patients who received a right-ventricular assist device (RVAD), the explant of that device was named to ``REXP".   For the visualization of the results, each type of AE and final outcomes were color coded, as shown in Fig. 3.
	
	\subsubsection{Forming patients' sequences}
	The sequence of AEs for each patient was identified and unified as a single record.  A sequence for $j_{th}$ patient ($P\textsubscript{j}$) can be presented as follow:

	\begin{equation}
	P_{j}: AE_1\rightarrow AE_2 \rightarrow..... \rightarrow   AE_i\\
	\end{equation}	
	Where:
	\begin{equation*}
	\begin{gathered}
	AE_1: \textrm{First adverse event}\\
	\mathrel{\makebox[\widthof{=}]{\vdots}} \\
	AE_i: \textrm{$i_{th}$ adverse event}\\
	\end{gathered}
	\end{equation*}
	
	As INTERMACS provided no information related to the order of concurrent AEs (negligible percent of AEs), they were alphabetically ordered in patients' sequences. 
	
	A consequence of the large number of event types (15) is the possibility of a large number of sparse patterns. For example, if we only considered sequences of length = 3, there would be 2,940 ($14 \times 14 \times 15$; ``Death" could be considered for only the last element) possible combinations. To overcome this issue, hierarchical clustering was used to divide the space of patients' sequences into more dense sub-spaces that represent patients with relatively similar sequences which would therefore be more amenable to pattern mining.
	
	\subsection{Hierarchical clustering}
	
	The goal of clustering is to identify clinically meaningful groups of patients with relatively similar sequences of post-LVAD AEs.

	\subsubsection{Defining the measure of dissimilarity}
	A distance matrix, \textbf{\textit{\textbf{d}}}, comprised of distances between each pair of patient sequences  was defined as: 
	
	\begin{equation}
	\begin{gathered}
	d (P_{n} ,P_{m})= {\vert{P_{n}}\vert} + {\vert{P_{m}}\vert} -2 {LCS(P_{n} ,P_{m})}\\
	\end{gathered}
	\end{equation}
	where, \textit{$\vert P_{n}\vert$} and \textit{$\vert P_{m}\vert$} are the lengths of the sequences for patients of \textit{$P_{n}$} and \textit{$P_{m}$}, respectively; \textit{$LCS(P_{n} ,P_{m})$} is the Longest Common Subsequence between \textit{$P_{n}$} and \textit{$P_{m}$}, as formulated below:
	
	
	\begin{equation}
	\begin{gathered}
	LCS(P_{n} ,P_{m})= max\{{{{\vert{l}}\vert}:l\in{SB(P_{n} ,P_{m})}}\}\\
	\end{gathered}
	\end{equation}
	
	Here, \textit{$SB(P_{n} ,P_{m})$} is a set of all common subsequences between \textit{$P_{n}$} and \textit{$P_{m}$} and\textit{ ${\vert{l}\vert}$} is the length of common subsequence.  A subsequence is a secondary sequence derived from another (primary) sequence  by deleting some or no elements while maintaining the same order of the remaining elements of the primary sequence. A \textit{common subsequence} is a subsequence that is common to both \textit{$P_{n}$} and \textit{$P_{m}$}. As an example, the only common subsequence between the \textit{$P_{1}$} , \textit{$P_{2}$}, shown in Fig. 4, is the subsequences of (Bleeding)-(Infection). Thus, in this example, the LCS is 2 and  \textit{d(\textit{$P_{1}$},\textit{$P_{2}$})} is 3 which means by  3 movements (deletions of respiratory failure and death from \textit{$P_{1}$}'s sequence and insertion a bleeding AE to \textit{$P_{1}$}'s sequence)  \textit{$P_{1}$}'s sequence becomes similar to  \textit{$P_{2}$}'s sequence.

	\begin{figure}[ht]
		\centering
		\includegraphics[width=0.4\textwidth,height=0.18\textheight]{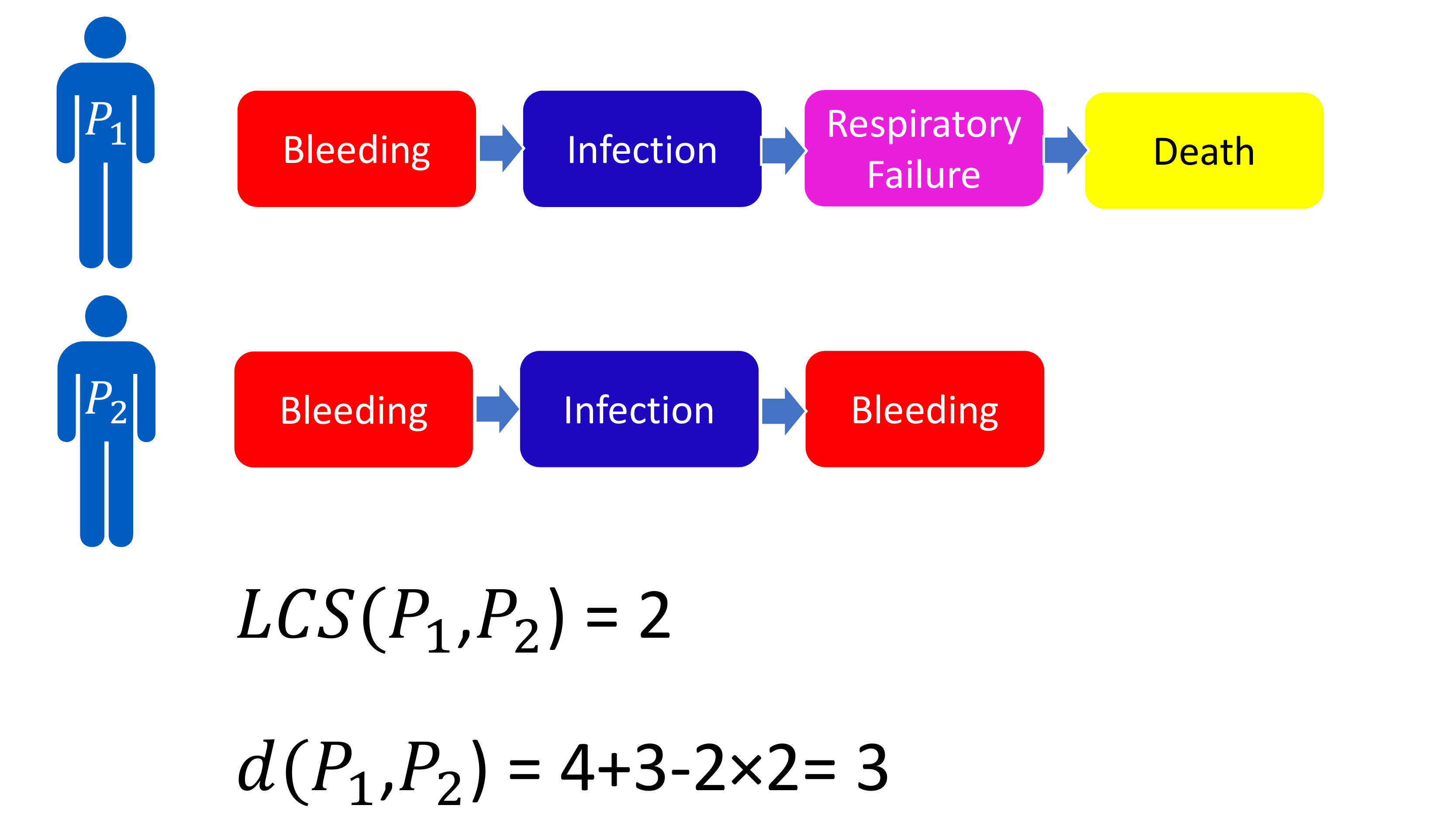}
		\caption{An example of computing dissimilarity score between two patients' sequences.}
	\end{figure}
	
	\subsubsection{Defining linkage method for hierarchical clustering}
	After forming dissimilarity matrix of AE sequences, they were  clustered using bottom to top hierarchical clustering with Ward linkage. Hierarchical clustering merges sequences with lowest distance, \textit{d} into a single group, and  updates the distance matrix for the  newly merged group and remaining patients. This merging process repeats using Ward's method in which groups of patients with lowest post-merging in-group variance (sum of squares) are merged until all patients are in one group. The Ward linkage distance between clusters is computed using Lance–Williams recurrence algorithm\cite{murtagh2014ward}. Briefly, considering two clusters of \textit{$C_i$} and \textit{$C_j$}, the distance between new cluster \textit{$C_{ij}$ (= $C_{i}$ $\cup$ $C_{j}$)} and remaining clusters such as \textit{$C_k$} is formulated as follows \cite{murtagh2014ward}:
	
	\begin{equation}
	\begin{gathered}
	D_{(ij,k)}= \alpha_i D_{(ik)}+ \alpha_j D_{(jk)}+ \beta D_{(ij)} + \gamma   {\vert{D_{(ik)} - D_{(jk)}}\vert} \\
	\end{gathered}
	\end{equation}
	Here, \textit{$D_{(ij,k)}$} is the distance between new cluster \textit{$C_{ij}$} and cluster \textit{$C_k$}. Lance–Williams coefficients  $\alpha$, $\beta$, and $\gamma$ are different for various linkage methods and are defined for Ward linkage as follows:
	
	\begin{equation}
	\begin{gathered}
	\alpha_z = \frac{\vert{z}\vert + \vert{k}\vert}{\vert{i}\vert + \vert{j}\vert + \vert{k}\vert} \: , \:z= i,j\\
	\beta = - \frac{\vert{k}\vert}{\vert{i}\vert + \vert{j}\vert + \vert{k}\vert}\\
	\gamma  = 0 \\
	\end{gathered}
	\end{equation}
	
	Here, $\vert {.} \vert$ indicates absolute value, and $i$, $j$, and $k$ are numbers of patients in each cluster.

	\subsubsection{Define criteria for choosing a number of clusters} 
	The clustering algorithm was implemented to maximize similarity between sequences within a group (internal validation), and minimize similarity between groups (external validation).

	\textit{Internal validation (the within-group similarity)}  was performed by extracting the most common subsequences and their \textit{support values}. This is the proportion of  sequences in a group that contain that specific subsequence and ranges from 0 to 1 (0\% to 100\%). The most common subsequences were extracted from AE sequences by applying a prefix-tree-based search algorithm using ``TraMineR" package from R\cite{masseglia2002algorithmes,grandstrand:2004}. It should be noted that this algorithm computes the support value for a given subsequence by including all longer sequences containing that subsequence. For instance, patients with the subsequence of (Infection)-(Bleeding) are counted among the patients with the subsequence of (Infection).

	\textit{External validation (the between-groups dissimilarity)} was performed by identifying the subsequences that best differentiate two groups of  patients' sequences using the Pearson Chi-square test (p-value of $\leq$ 0.01).

	\textit{Step-wise evaluation of clustering:} 
	The clustering evaluation for choosing the number of groups (n) started with evaluating the two-cluster solution (n= 2) and evaluation was continued for bigger numbers of groups until both internal and external criteria were satisfied for all the groups. At each step of clustering evaluation, n was increased by 1; only one group was divided into two new groups (G1 \& G2) and their qualities (high internal similarity and low external similarity) were evaluated. First, external validation was checked between the G1 and G2. If the external criteria were satisfied, internal validations will be checked for each of the G1 and G2. If any of  G1 or G2 satisfied the internal validation, it was considered as a qualified group, otherwise it  was considered as an unqualified group that needed to be split into  sub-groups with more similar sequences in the later steps of clustering evaluation (bigger values of n). 
	
	\textit{Interactive visualization evaluation:} 
	To help visualize the composition of AEs within each group (or sub-group), a histogram was constructed using the same color coding from Fig. 3. in which the proportions of each category of AE was plotted for each position in the sequence. Fig. 5 provides the histogram for the aggregate of all  the 13,192 sequences of AEs  over their chronological positions in the sequences. The first column of the graph, AE\textsubscript{1}, presents the proportions of various types of AE that  patients experienced as their first AEs (the first element of the sequences of AEs), the second column for the second AEs, and so forth, through the thirty sixth AE. The uniformity of the distribution in the first set of columns of the histogram (e.g. AE\textsubscript{1} through AE\textsubscript{20}) is contrasted with the heterogeneity of the subsequent columns. This reflects the increasing diversity of AE's as there is a decreasing number of patients with longer sequences. For instance,  there is only one patient with a sequence of 36 AEs - the last AE being death (yellow). This type of visualization is helpful to get a general quick view of the distribution of AEs in a cluster of patients.
	
	\begin{figure}[h]
		\centering
		\includegraphics[width=0.3\textwidth,height=0.18\textheight]{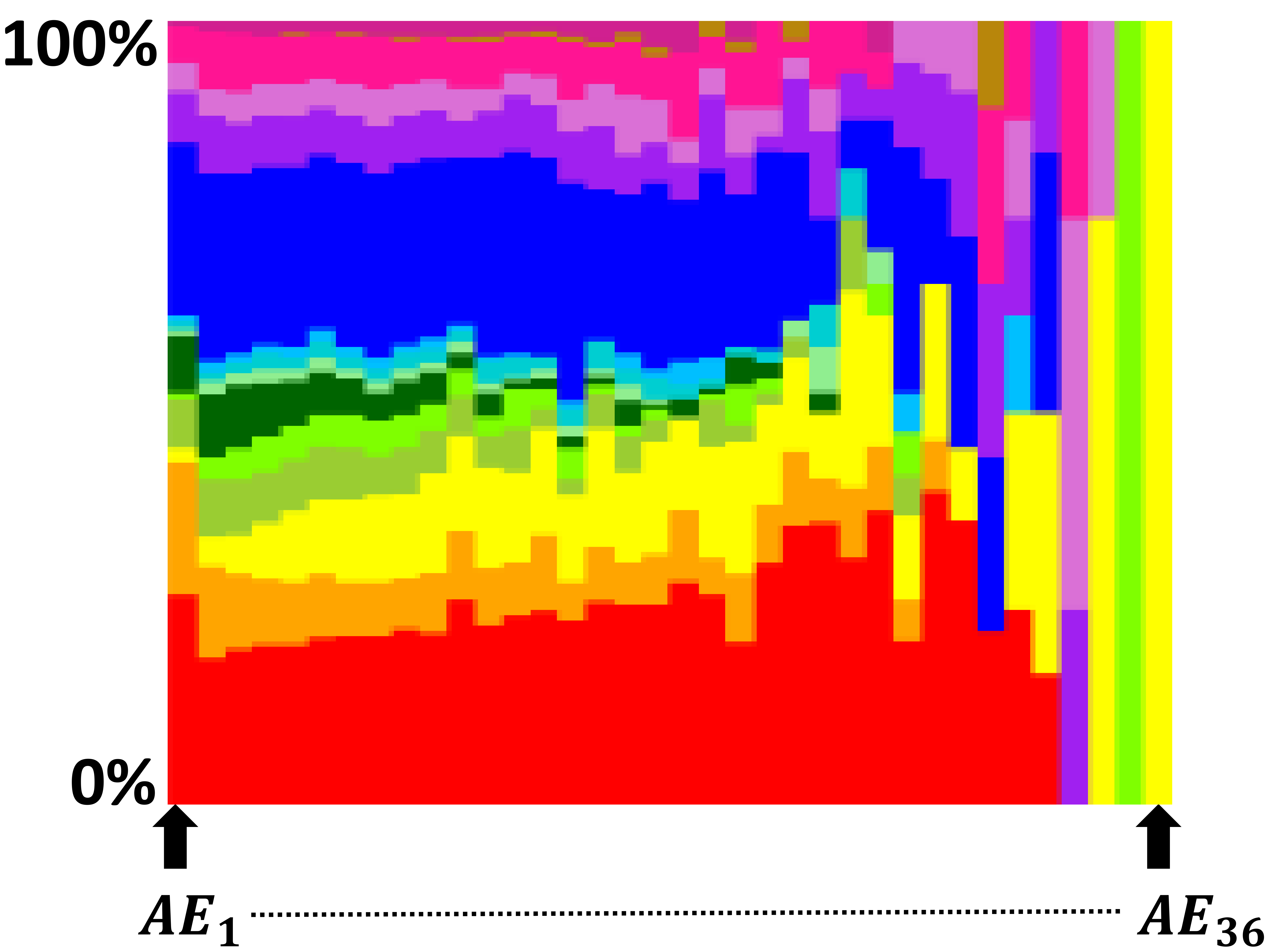}
		\caption{The proportions of various types of AE in all the 13,192 sequences of AEs in this study for temporally ordered AEs of patients sequences (see Fig. 3. for color coding).}
	\end{figure}
	
	The final clusters  were evaluated by our clinical experts to determine the reasonableness of the clusters.

	\subsection{Markov Chain Models of AEs} 
	Following clinical confirmation of the results of hierarchical clustering, patterns of AEs for each group (cluster) were analyzed using Markov modeling (MM). This has been shown to be  a useful tool to model repetitive events, where the timing and the order of events are important \cite{sonnenberg1993markov}. For instance, the transition of bleeding  to infection can be considered different  if bleeding occurred as the 1\textsuperscript{st} AE followed immediately by  infection, (Bleeding1)-(Infection2), versus bleeding occurred as the 2\textsuperscript{nd} AE,  and then infection occurred,  (Bleeding2)-(Infection3). Accordingly, the transitions between sequential AEs were assessed for likelihood of transitions as a function of the chronological position in the sequence of AEs. As this is the first attempt to model the sequential AEs after LVAD implant, it was preferred to start with a simple solid model such as the first-order Markov chain. MM was defined for a sequence of $AE_1$, $AE_2$, $AE_3$,... for discrete points of time: $1,...,n$ was defined as follows:
	
	\begin{equation}
	\begin{gathered}
	P( AE_{n+1}=ae \mid AE_1 = ae_1 ,AE_2 = ae_2,....,AE_n = ae_n)\\
	= \\
	P(AE_{n+1}=ae \mid AE_n = ae_n )
	\end{gathered}
	\end{equation}
	\noindent
	where $ae$ was an AE from 15 different types of AE that could occur in various orders in
	a sequence of AEs. The above formula assumes that the probability of transitioning to the next state, $AE_{t+1}$, depends only on the present state, $AE_{t}$. Another assumption of MM is homogeneity, which assumes all patients in the same state have the same risk of transition. 	
	
	MM considers each types of AE as unique Markov \textit{state}, and transitions between states as \textit{events}. As an example, a sequence of (Bleeding1)-(Infection2)-(Death3) was represented as (Bleeding1 $\rightarrow$ Infection2) $\Rightarrow$ (Infection2 $\rightarrow$ Death3) that includes 2 events presenting 2 transitions between 3 states/AEs in the sequence. It also shows the transitivity relation between 2 events as the target state in the first event, Infection2,  is the source state in the second event. A transition matrix was formed by computing transition probabilities between each pair of states/AEs from all the sequences of AEs in each group. Then, chains of events were extracted from the transition matrix by connecting events that have transitivity relations. Finally, thresholding of the extracted chains was performed,  based on the distributions of  transition probabilities and frequency of occurrence to eliminate the ``noise" of numerous infrequent or rare transitions. The thresholds were chosen subjectively as 0.1  for transition probability and between 30 to 50 for frequency of occurrences, respectively,  to achieve a compromise between reducing noise versus over-simplifying the resulting collection of MMs. 
	
	\section{Results}
	This study was performed with data from INTERMACS for 13,192 patients with advanced heart failure who underwent continuous flow LVAD implant between 2006 and 2015. A total number of 58,575 AEs, including 15 various types of AE, were included in this study. Table. \rom{1} summarizes the 15 types AEs and final outcomes, and indicates that ``Bleeding", ``Infection",  and ``Cardiac Arrhythmia" are the most common AEs.
	
	The time of recorded AEs ranged between 0 (at the time of implant) and 87 months (7.25 years) after LVAD implant with mean of 9.95 months.  A great proportion of AEs (81\% of total AEs) occurred within the first 18 months after LVAD with the peak of AEs (27\%) during the first month.

	\begin{table}[h]\centering
		\renewcommand{\arraystretch}{1.3}
		\caption{Frequency of Various Types of Events, Including AEs and Final Outcomes, and Their Percentages of the Total Number of Events. The Cells in the Table That Contains the Final Outcomes Including ``Death", ``Explant", and ``Explant: Transplant" Are Highlighted in Light Orange Color. }
		\begin{tabular}{l|cc}
			\multicolumn{1}{c}{Event} & \multicolumn{1}{c}{Frequency} & \multicolumn{1}{c}{Percent}\\	
			\midrule	
			Bleeding  & 12,877 & 22.0\\ \hlineB{2}
			Cardiac Arrhythmia & 6,361 & 10.9   \\ \hlineB{2}
			Device Malfunction & 3,726 & 6.3  \\ \hlineB{2}	
			Hemolysis & 797 &  1.4 \\\hlineB{2}
			Hepatic Dysfunction &  850 & 1.4  \\ \hlineB{2}
			Hypertension & 701 & 1.2  \\\hlineB{2}
			Infection & 13,399 & 22.9 \\\hlineB{2}
			Neurological Dysfunction & 3,873 &  6.6 \\ \hlineB{2}
			Renal Dysfunction & 2,248 &  3.8 \\ \hlineB{2}
			Respiratory Failure & 3,614 & 6.2  \\ \hlineB{2}
			REXP (RVAD explant) & 92  &  0.1 \\\hlineB{2}	
			Right Heart Failure & 747 &  1.3 \\ 
			\noalign{\global\arrayrulewidth=1mm}
			\arrayrulecolor{black}\hline
			\noalign{\global\arrayrulewidth=0.2mm}
			
			\cellcolor{orange!30}Death &  3,675 &  6.3 \\ \hlineB{2}
			\cellcolor{orange!30}Explant & 1,819 &  3.1 \\ \hlineB{2}
			\cellcolor{orange!30}Explant: Transplant & 3,796 &  6.5 \\ 
			
			\noalign{\global\arrayrulewidth=1mm}
			\arrayrulecolor{black}\hline
			\noalign{\global\arrayrulewidth=0.2mm}
			
			\rowcolor{gray!30}Total recorded events  & 58,575 &  100.0 \\ \hlineB{2}
		\end{tabular}
		

	\end{table}
	
	\subsection{Sequences of AEs} 
	
	The length of AE sequences ranged from 1 to 36, however 94\% of the sequences were less than or equal to 10. The distribution of lengths $\leq$15 are provided in Fig. 6. (Lengths 16-32 were very rare, $<$1\%, and therefore not shown).

	\begin{figure}[h]
		\centering
		\includegraphics[width=0.49\textwidth,height=0.15\textheight]{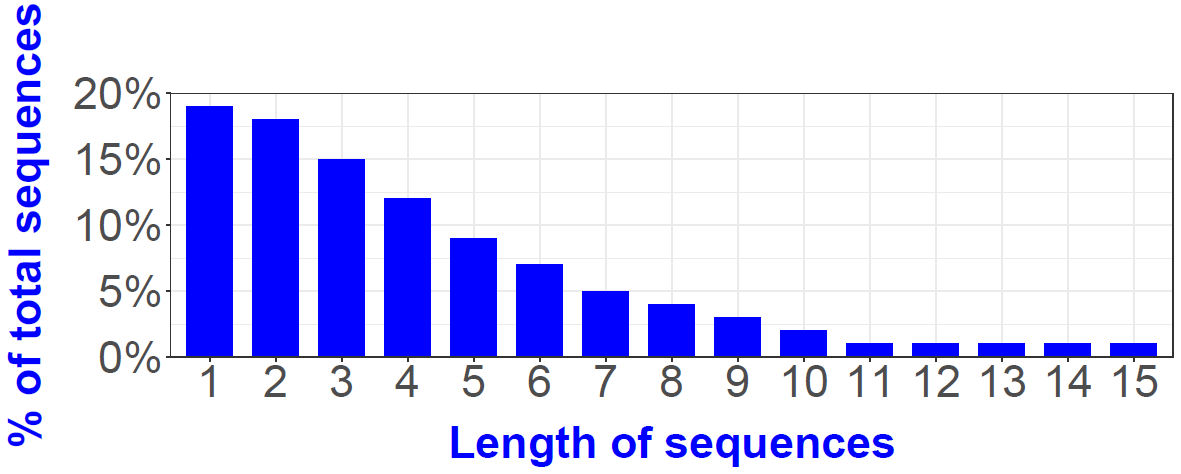}
		\caption{length of AE sequences from 1 to 15. (Length of 16-32, not shown, were rare $<$ 1\%.}
	\end{figure}
	
	
	\subsection{Hierarchical Clustering }
	
	Fig. 7 shows the distribution of dissimilarity scores between AE sequences, which is left-skewed. The min/max of scores were 0/53 with the mean of 10 and the median of 8. The numbers of dissimilarity scores less than 2 and greater than 22 were negligible ($\approx$0\%).
	
	\begin{figure}[h]
		\centering
		\includegraphics[width=0.35\textwidth,height=0.15\textheight]{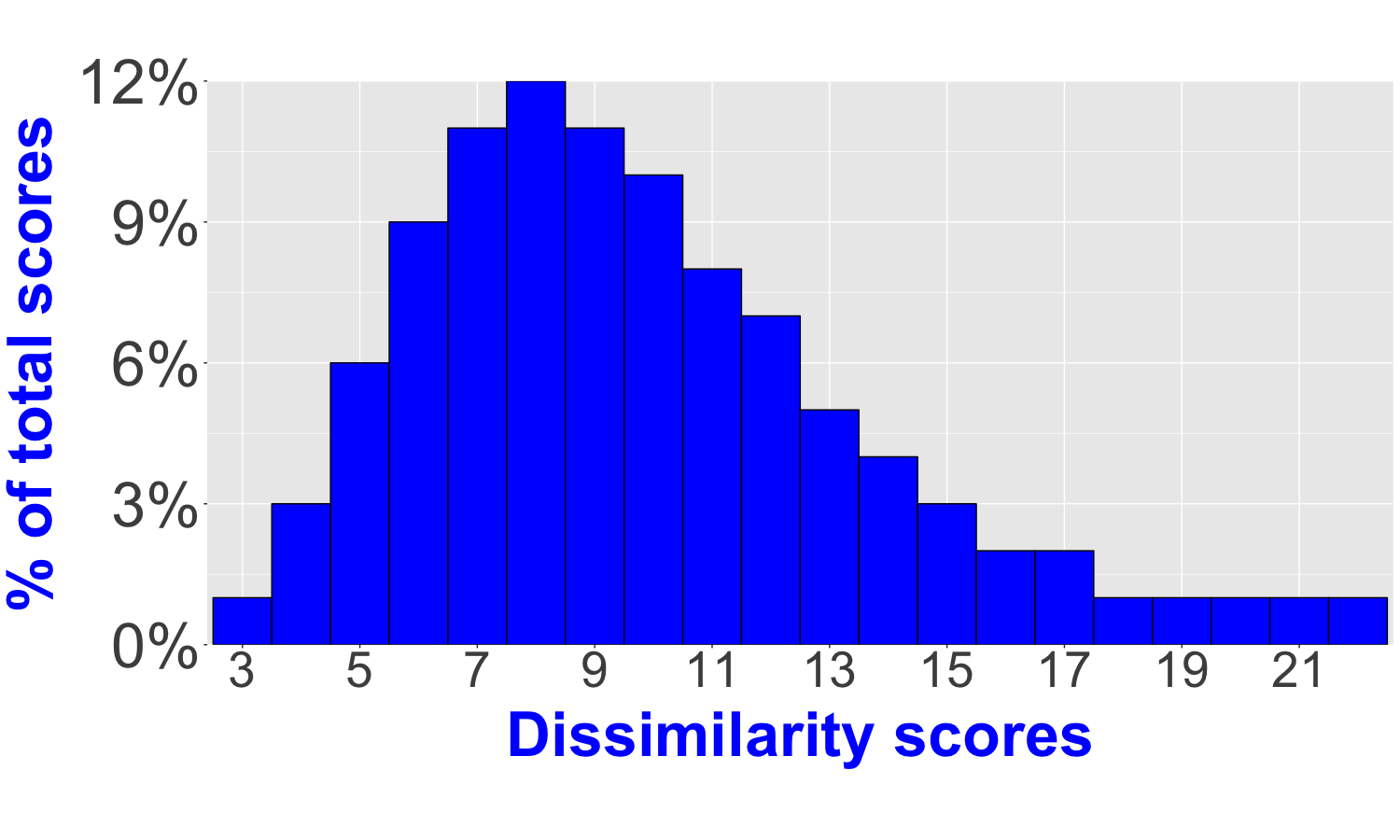}
		\caption{The dissimilarity scores distribution of all the sequences of AEs}
	\end{figure}

	\begin{figure*}[]
		\begin{subfigure}{0.5\textwidth}\centering
			\centering
			\includegraphics[width=0.9\textwidth,height=0.25\textheight]{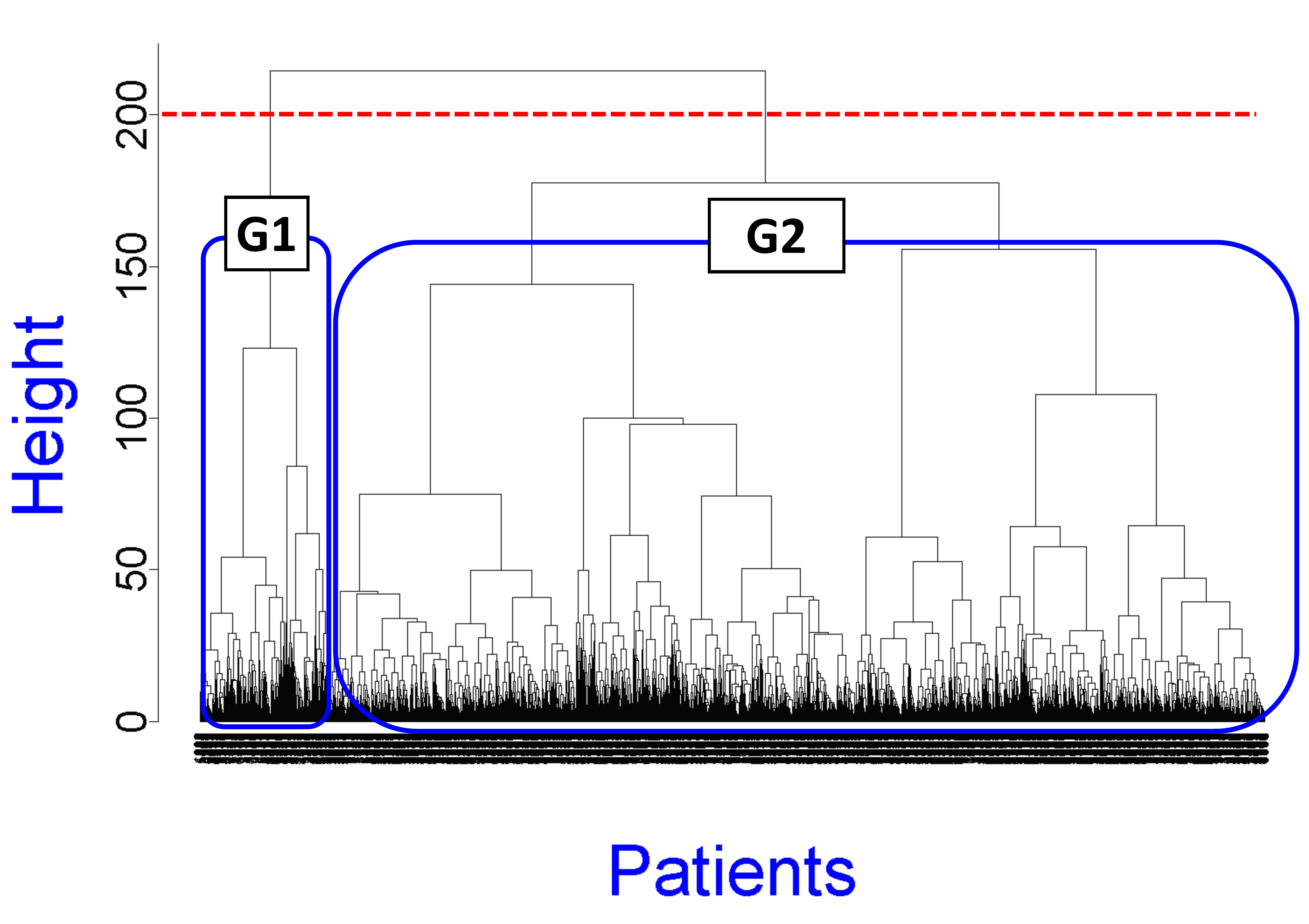}
			\caption{ Cutting the dendrogram into the two clusters.}
		\end{subfigure}
		\begin{subfigure}{0.5\textwidth}\centering
			\centering
			\includegraphics[width=0.9\textwidth,height=0.25\textheight]{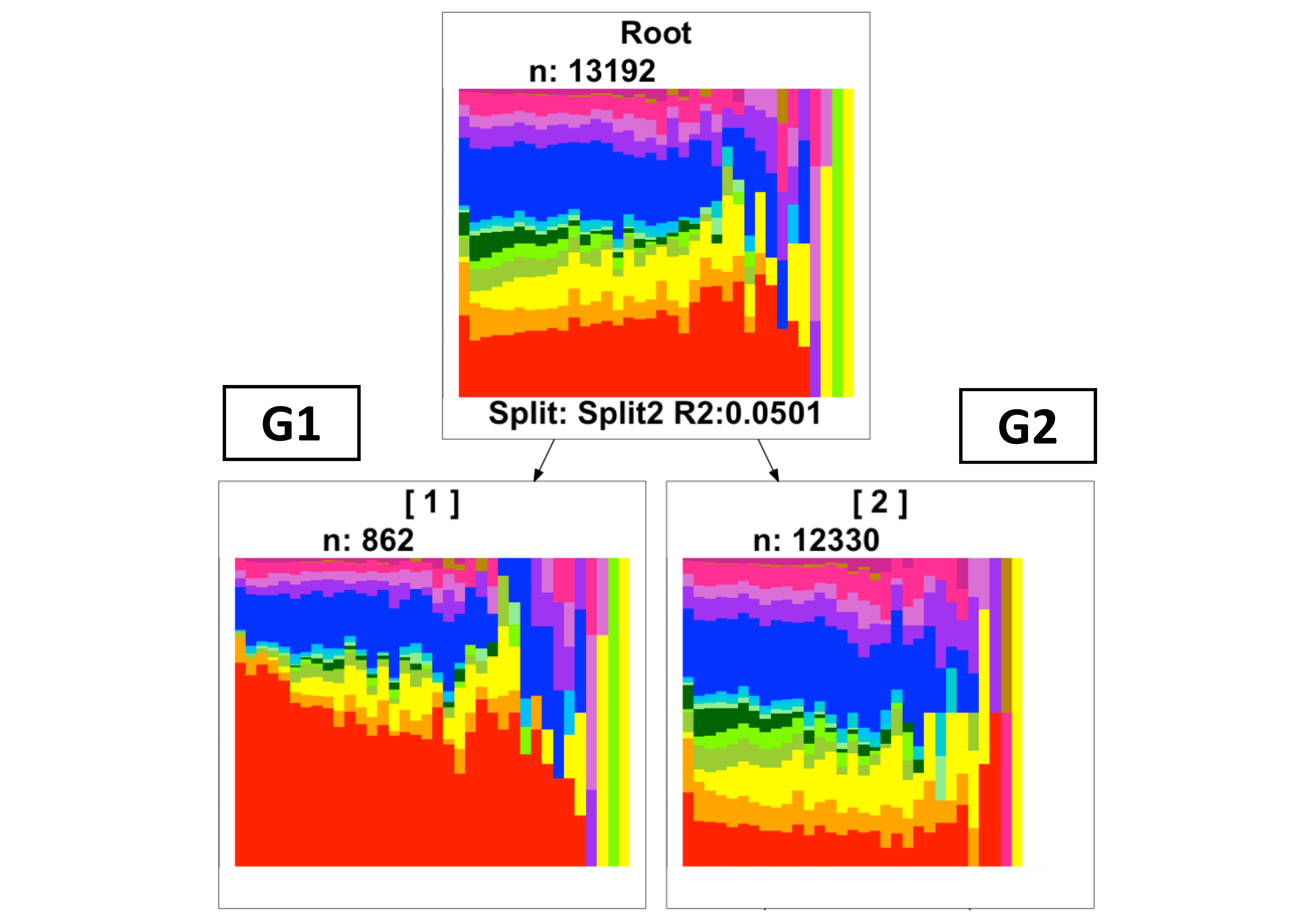}
			\caption{A regression tree plotting distributions of various types of AE over temporally ordered AEs in patients' sequences.}
		\end{subfigure}
		\begin{subfigure}{0.5\textwidth}\centering
			\centering
			\includegraphics[width=1\textwidth,height=0.25\textheight]{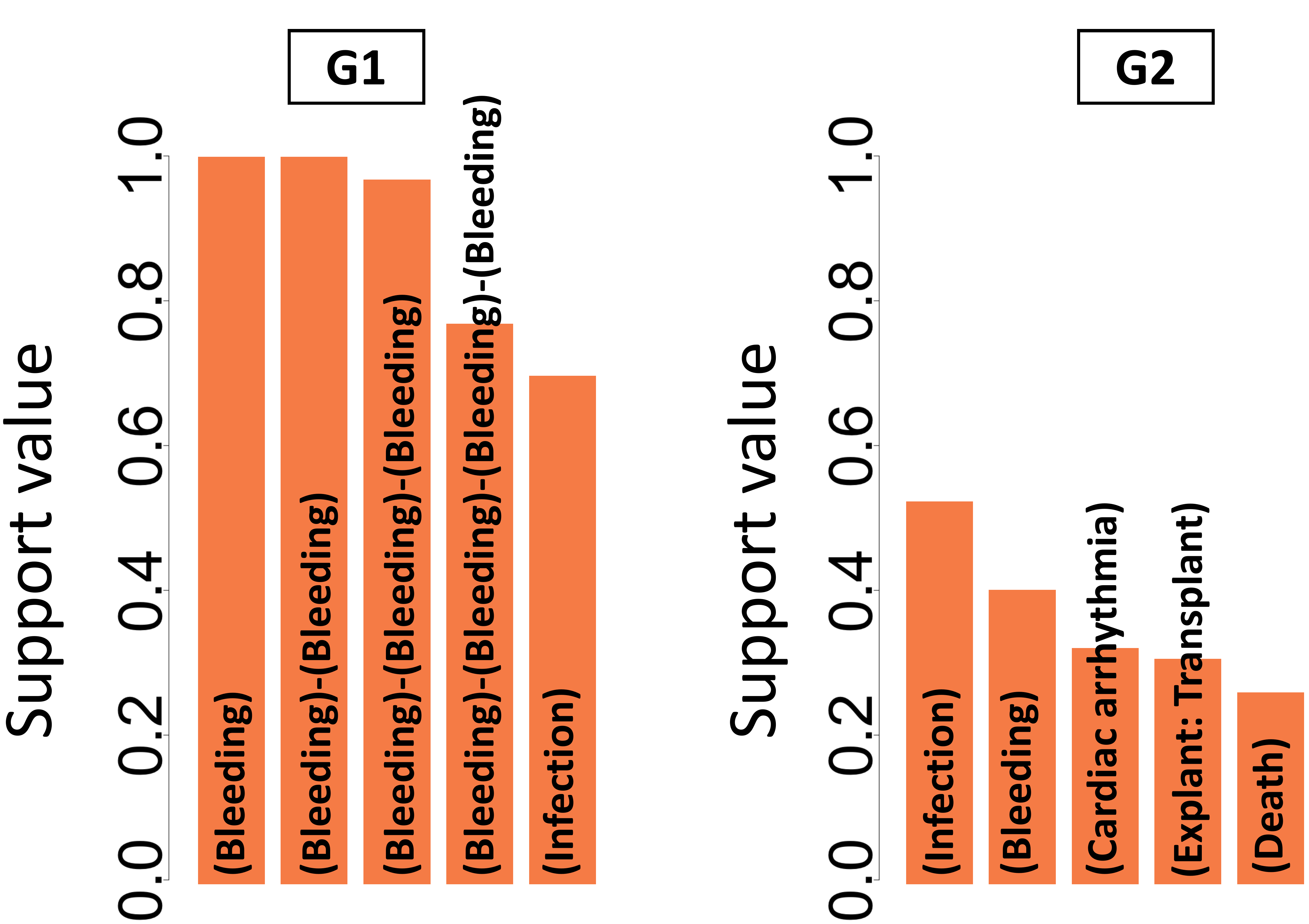}
			\caption{Internal validation: support numbers for most common subsequences in two new-formed clusters.}
		\end{subfigure}
		\begin{subfigure}{0.5\textwidth}\centering
			\centering
			\includegraphics[width=0.85\textwidth,height=0.09\textheight]{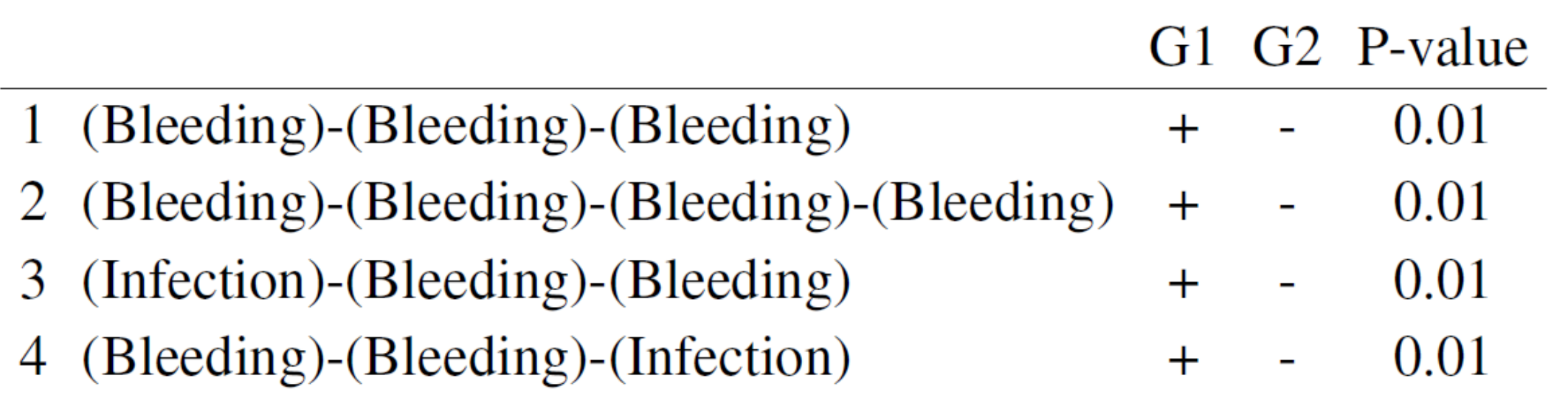}
			\captionsetup{width=0.8\linewidth}%
			\caption{External validation: the first four discriminative subsequences between G1 and G2. +/- determines the sign of Pearson's residual.}
			
		\end{subfigure}
		\caption{First step of step-wise cluster evaluation: evaluation of the two-cluster solution}
	\end{figure*}

	The result of hierarchical clustering is presented as a dendrogram, shown in Fig. 8(a), which depicts the taxonomic relationship between clusters formed at each level of grouping. The bottom of the dendrogram represents all the patients’ dissimilarity scores, which is a dense dark area because of the large number of patients. Moving from bottom to top of the dendrogram, one cluster is formed at each level of hierarchical clustering by grouping two sub-clusters until, at the top of the dendrogram, all patients are in a single group. The y axis, “Height”, represents the dissimilarity between clusters (groups of patients) at that level of the tree, which is measured using Ward linkage.
	
	The eventual partitioning of the dendrogram was based on the three criteria: the internal validation (within-group similarity), external validation (between-groups dissimilarity), and clinical interpretation. This was performed iteratively by “cutting” the dendrogram horizontally. An example is shown in Fig. 8(a) in which a horizontal cut results in two groups (G1 and G2). The first group includes 862 patients (7\% of total patients), and the second group that includes 12,330 patients (93\%). Their corresponding histograms (similar to Fig. 5) are provided in Fig. 8(b), in which the proportions of various types of AE are temporally ordered. Visual inspection of the histogram corresponding to G1 reveals an obvious dominance of the Bleeding AE (red color). This also implies a degree of similarity of patients in this group. By contrast, the histogram corresponding to G2 does not present any single dominant AE. This is understandable since this group is comprised of a much larger proportion of the total number of patients (93\%). Thus, the patients in the second group are not similar and require further stratification.

	Fig. 8(c) represents the support values of the five most common subsequences in these two groups. In the first group (G1), the unary sequence (Bleeding) and the binary sequence (Bleeding)-(Bleeding) are the most common, both with support value of 1, indicating that 100\% of patients in this group had a minimum of two bleeding AEs in their sequences. The next most common sequences were found to be (Bleeding)-(Bleeding)-(Bleeding) and (Bleeding)-(Bleeding)-(Bleeding)-(Bleeding), with support values of 0.98 and 0.73. It was concluded that the sequences in the first group all shared common subsequences, indicating high within-group similarity. In contrast, the maximum support value for the second group (G2) was approximately 50\% corresponding to the (Infection) subsequence. This indicated low similarity within this group, and hence does not pass the internal validation.
	
	As a final test, Fig. 8(d) shows the external validation in which subsequences which best discriminate sequences of the two groups via Pearson Chi-square test. The table shows the first four most discriminative subsequences, in which the plus (+) and minus (-) sign indicates that the observed frequencies of the subsequences were higher (+) or lower (-)	than equally distributed frequencies. Here, p-values below 0.05 were taken to indicate discrimination between groups. The subsequence of (Bleeding)-(Bleeding)-(Bleeding) was the most discriminating subsequence with p-value of 0.01 and support percentage for G1 greater than 98\%. All of the remaining discriminative subsequences consisted of at least two bleeding AEs, emphasizing that multiple bleeding AEs was responsible for differentiating G1 from G2. Accordingly, it was concluded that the groupings were externally validated.

	In summary, the initial two-cluster solution passed the external validation, but only G1 passed the internal validation. However, the second group which includes the remaining 93\% of patients required further subdivision. This was accomplished in a similar manner, by bisecting the dendrogram at a lower level, effectively separating the second group into two sub-groups of 6,168 and 6,162 patients. This was followed by the same process to evaluate the external and internal validation. This procedure was repeated until all the resulting groups passed validation, resulting in a final number of seven groups (summarized in Table. \rom{2}). For convenience, each of the seven groups was given a mnemonic name based on visual inspection of the histogram including: GRP1: ``Recurrent bleeding", GRP2: `` Trajectory of device malfunction \& explant", GRP3: `` Infection", GRP4: ``Trajectories to transplant", GRP5: `` Cardiac arrhythmia", GRP6: ``Trajectory of neurological dysfunction \& death", and GRP7: `` Trajectory of respiratory failure \& renal dysfunction \& death". For example, this histogram of GRP1 reveals an obvious dominance of the Bleeding AE (red color), and was therefore given the name “Recurrent Bleeding.” In a similar fashion, G2 revealed a dominance device malfunction (forest green) and explant (lime green.) Since the two are always related sequentially, this group was named “Trajectory of device malfunction \& explant.”


	Fig. 9 demonstrates clustering results through a regression tree. There is histogram associated with each of the groups; the ordinate of which reflecting the proportion of each color-coded AE type (from 0\% to 100\%) and abscissa reflecting the location in the respective sequences. Since each group has a unique maximum length, this is reflected in the varied width of these plots. Each of these groups were assigned a descriptive title that reflected the dominant AE or AEs therein. For instance, the dominant colors in the GRP2 plot are yellow green (representing device malfunction) and spurious  green (representing Explant outcome). Sequence analysis for GRP2, similar to sequences analysis for GRP1 in Fig. 8(c)\&(d), reveals that 1,097
	out of 1,193 patients who experienced device malfunction
	eventually had the device explanted, indicating the temporal pattern (subsequence) of (Device Malfunction)-(Explant). The sequence analysis was performed for each of the seven groups.



	\begin{figure}[]
		\centering
		\includegraphics[width=0.35\textwidth,height=0.5\textheight]{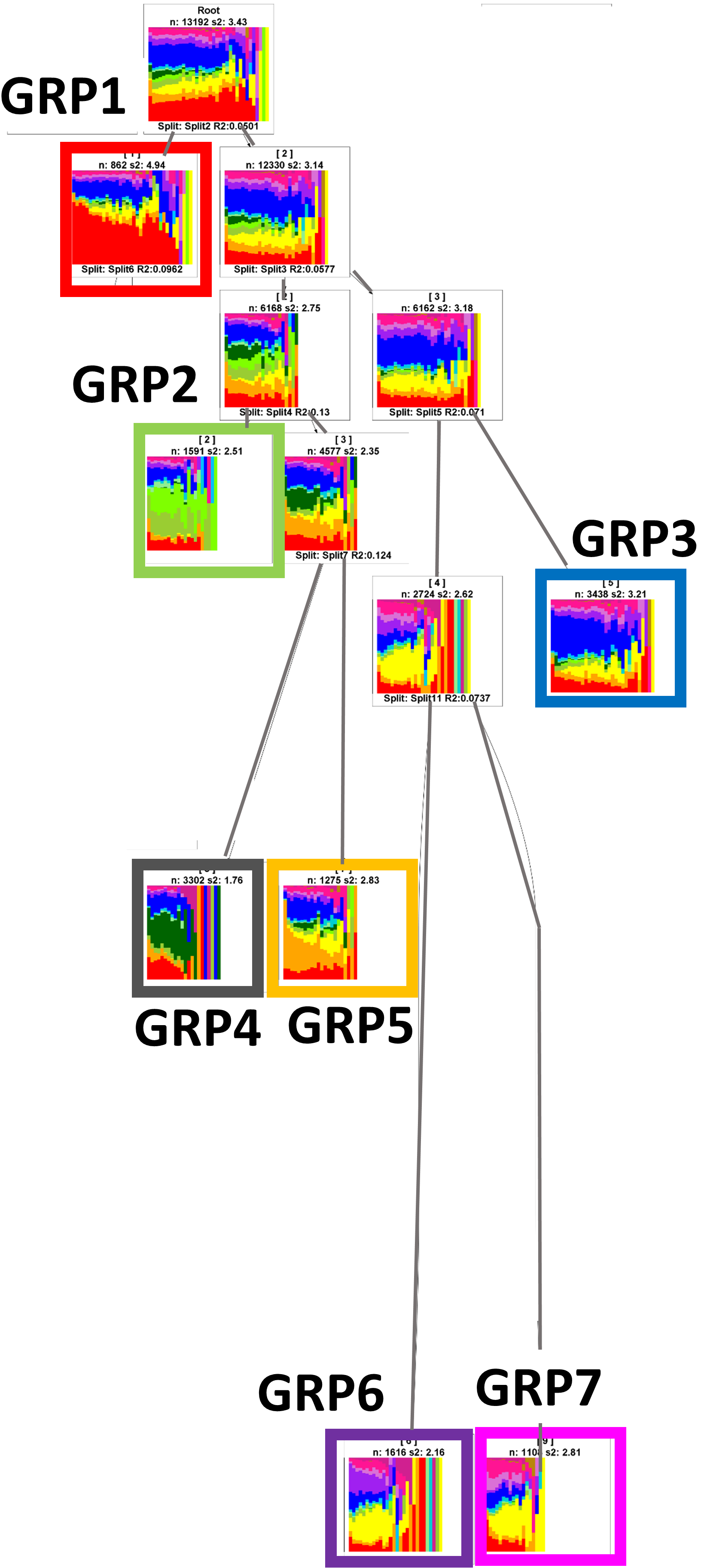}\\
		
		\caption{Regression tree showing groups formed at each step of hierarchal clustering. Groups are split to smaller groups until both internal and external criteria are satisfied. Groups are numbered based on the steps in which they are formed.}
	\end{figure}			
	
	\begin{table*}[h]\centering
		\caption{Summary of Statistical Information Related to Groups Resulted from Hierarchical Clustering.}
		\renewcommand{\arraystretch}{1.5}
		\footnotesize
		\begin{tabular}{cc|ccc|cc|cc}
			
			&   & \multicolumn{3}{c}{Number of AEs} & \multicolumn{2}{c}{AEs post-LVAD time (month) \textsuperscript{$\dagger$}} & \multicolumn{2}{c}{AEs time span (month) \textsuperscript{$\dagger\dagger$}}  \\
			
			Number of patients (*) & Number of AEs (**) &   {Min/Max} & {Mean} & {Median}   & {Mean} & {Median} & {Mean} & {Median} \\ \hline\hline
			
			\rowcolor{gray!10}
			\multicolumn{9}{c}{\textcolor{red}{\textbf{GRP1:Recurrent bleeding}}} \\ \hline  
			862 ($\approx$7\%) & 
			8,388 ($\approx$14\%) &
			
			3/36 &
			9.73 &
			9 &
			12.55 &
			7 &
			19.48 &
			16
			\\\hline

			\rowcolor{gray!10}
			\multicolumn{9}{c}{\textcolor{green}{\textbf{GRP2:Trajectory of device malfunction \& explant}}} \\ \hline
			
			1,591 ($\approx$12\%)  & 
			7,049 ($\approx$12\%) &
			
			1/21 &
			4.43 &
			4 &
			11.19 &
			6 &
			9.89 &
			4
			\\\hline

			\rowcolor{gray!10}
			\multicolumn{9}{c}{\textcolor{blue}{\textbf{GRP3:Infection}}} \\ \hline
			3,438 ($\approx$26\%)    &
			15,771 ($\approx$27\%)   &
			
			1/31 &
			4.59 &
			3 &
			11.31 &
			6 &
			10.25 &
			4
			\\\hline

			\rowcolor{gray!10}
			\multicolumn{9}{c}{\textcolor{OliveGreen}{\textbf{GRP4:Trajectories to heart transplant}}}\\ \hline
			3,302 ($\approx$25\%)    & 
			10,093 ($\approx$17\%)  &

			1/22 &
			3.06 &
			3 &
			7.98 &
			5 &
			7.31 &
			4
			\\\hline

			\rowcolor{gray!10}
			\multicolumn{9}{c}{\textcolor{orange}{\textbf{GRP5:Cardiac arrhythmia}}}\\ \hline
			1,275 ($\approx$10\%)  & 
			5,715 ($\approx$10\%)  &

			1/22 &
			4.48 &
			3 &
			8.63 &
			3 &
			9.48 &
			4
			\\\hline

			\rowcolor{gray!10}
			\multicolumn{9}{c}{\textcolor{Plum}{\textbf{GRP6:Neurological dysfunction \& death}}}  \\ \hline
			
			1,616 ($\approx$12\%) & 
			5,911 ($\approx$10\%)   &
			
			1/28 &
			3.66 &
			3 &
			9.55 &
			5 &
			6.42 &
			1
			\\\hline

			\rowcolor{gray!10}
			\multicolumn{9}{c}{\textcolor{Magenta}{\textbf{GRP7:Trajectory of respiratory failure \& renal dysfunction \& death}}}\\ \hline
			
			1,108 ($\approx$8\%) & 
			5,648 ($\approx$10\%) &
			
			1/18 &
			5.10 &
			5 &
			6.06 &
			1 &
			5.78 &
			1
			\\\hline
			
		\end{tabular}
		\bigskip
		
		\small\textsuperscript{*} \% of the total 13,192 patients in this study\\
		
		\small\textsuperscript{**} \% of the total 58,575 recorded AEs  in this study\\
		
		\small\textsuperscript{$\dagger$}AEs post-LVAD time is based on the month after LVAD implant (0 post-LVAD month means AE occurred at the time of LVAD implant) \\
		
		\small\textsuperscript{$\dagger\dagger$} AEs time span = time of the last AE (post-LVAD month) - time of first AE (post-LVAD month) \\

	\end{table*}

	Table. \rom{2} provides statistics of each patient's group. GRP3 and GRP4 had the highest number of patients and AEs by having 26\% and 25\% of total number of patients, respectively, and 27\% and 17\% of total recorded AEs, respectively, in this study. On the other hand, GRP1 had the lowest number of patients, 7\% of the total number of patient (862 patients), while they had 14\% of total AEs. In addition, patients in GRP1 had the greatest number of AEs with average number of 9.73 AEs, and minimum and maximum numbers of 3 and 36 AEs. The average numbers of AEs in other groups were  $\leq$5 and minimum numbers of AEs were 1. Columns of 6 and 7  of Table. \rom{2} shows information related to the time of AEs occurrences measured by the months after the LVAD implants. The distributions of post-LVAD time (month) of AEs occurrences were skewed to the right in all the groups as the means were greater than the medians. The average time of AEs occurrences were less than 13 months in all the groups and the median time were less than 7 months. AEs of GRP7 occurred  at the earliest post-LVAD time by average of 6.06 months after LVAD  and median of 1 month after LVAD. The last two columns of Table. \rom{2} presents information related to the time span  of patients' AEs (time of last AE - time of first AE) measured in month. The distributions of time span  of patients' AEs were also skewed to right indicating higher number of patients experienced AEs in a short time span. GRP1 had the longest time span of AEs by average of 19.48 months between the first AE and last AE, while, the GRP6 and the GRP7 had the shortest time span by average of approximately 6 months and median of 1 month.
	
	\begin{figure*}[]	
		\centering
		\includegraphics[width=0.9\textwidth,height=0.07\textheight]{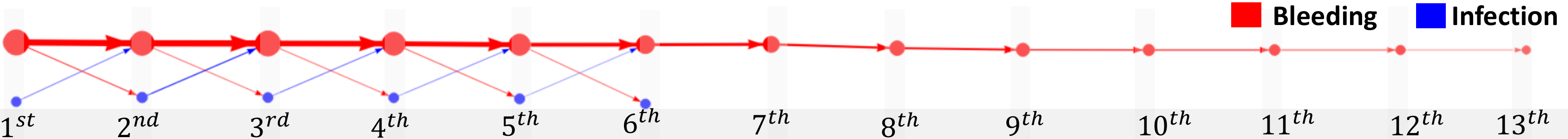}
		\caption{MM of GRP1: Recurrent bleeding (n= 862 patients)}
	\end{figure*}

	\subsection{Markov Chain Models of AEs}
	The following sections presents results of Markov modeling (MM) within each of the groupings of patients presented above. The chains of transitions between AEs are presented with the graphs in which the size of the circles represent the frequency of AEs at each position in the chain and the thickness of the arrows reflecting the frequency of each AE that are followed by the subsequent AE. 

	\begin{itemize}[leftmargin=*]	
		\item \textbf{\textit{GRP1: Recurrent bleeding}}
		
		All the 862 patients in GRP1 had at least two bleeding AEs, and among them, 98\% (847 patients) had at least three bleeding AEs. The Markov chain for GRP1, shown in Fig. 10 is characterized by a long sequence of recurrent bleeding AE’s (red circles), with a limited amount of branching involving an intermediate infection AE (blue circles). The frequencies of transitions depicted by the thickness of the arrows is seen to diminish progressively along the chain over the time. For example, the transition frequency for the 1\textsuperscript{st} through 5\textsuperscript{th} bleeding exceeded 300; but beyond the 5\textsuperscript{th} bleeding event, the frequency reduced to the range 100-250. This analysis revealed that the probability of recurrent bleeding exceeded 50\% for all instances up to the 13th AE. The transitions probabilities from infection AEs to bleeding AEs were 50\% to 65\%, while from bleeding AEs to infection AEs were 14\% to 18\%.

		\item \textbf{\textit{GRP2: Trajectory of device malfunction \& explant. }}
		
		\begin{figure}[h]
			\centering
			\includegraphics[width=0.4\textwidth,height=0.2\textheight]{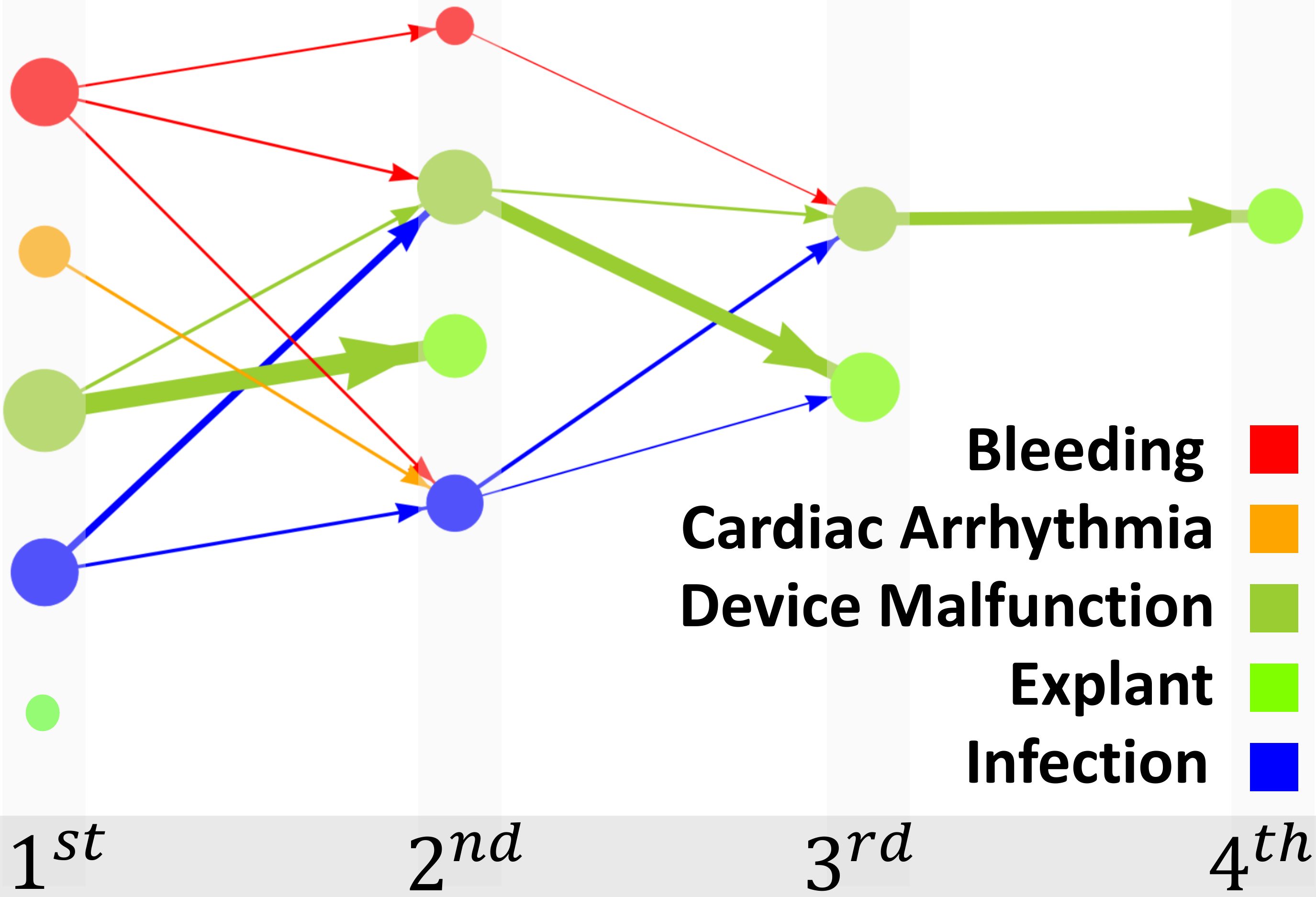}
			\caption{MM of GRP2: Trajectory of device malfunction \& explant (n= 1,591 patients)}
		\end{figure}
	
	\begin{figure*}[]	
		\centering
		\includegraphics[width=0.9\textwidth,height=0.12\textheight]{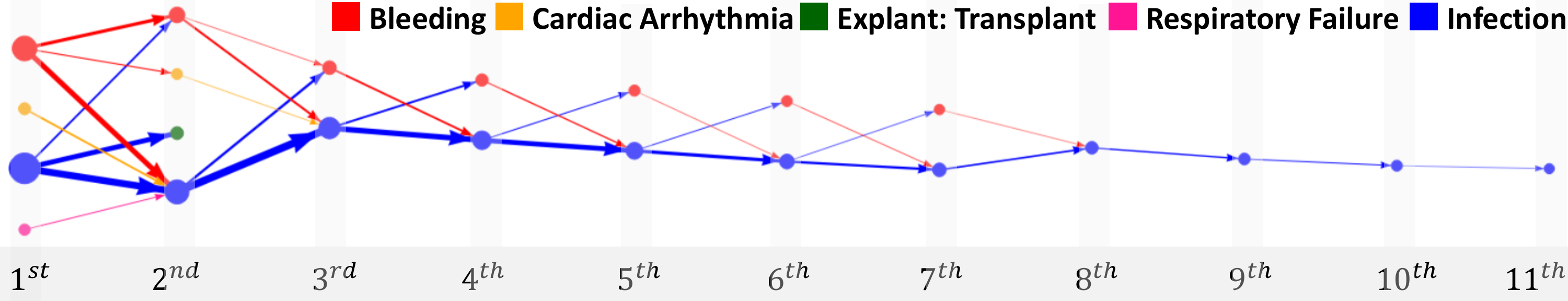}
		\caption{MM of GRP3: Infection (n= 3,438 patients) }
	\end{figure*}
		
		75\% of 1,591 patients in GRP2 experienced a device malfunction and 94\% had Explant as their final outcome. The Markov chain for GRP2 (Fig. 11) was found to be much more diverse than the chain for GRP1. Multiple paths involving device malfunction (dark green) were found, although the terminal event was most commonly ($>$50\%) device explant (light green). Only a small number of patients in this group (n=111, approximately 7\%) experienced device explant as the initial, isolated AE (indicated by the small light green node at the bottom of the 1\textsuperscript{st} column of Fig. 11). The majority of patients for which device explant was recorded was preceded by another AE, most commonly bleeding and infection (probability between 20-30\%). There were also about 18\% and 19\% probabilities of recurrent device malfunction as the 2\textsuperscript{nd} AEs or the 3\textsuperscript{th}  AEs, respectively. The frequency of transition from Infection as the 2\textsuperscript{nd} AE to Explant was 21\%, but with no reported device malfunction AE (n=211, the spurious green circle in the 1\textsuperscript{st} column).

		\item \textbf{ \textit{GRP3: Infection}}
		
		The Markov chain for GRP3 (Fig. 12) is characterized by a long sequence of infection AEs (blue circles), corresponding to a total number
		of 6,462 recorded infection AEs for 3,438 patients in GRP3, with a limited amount of branching involving an intermediate bleeding AE (red circles). This analysis revealed that the probability of recurrent infection ranged from 34\% to 49\%. The frequencies of recurrent infection AEs depicted by the thickness of the arrows is seen to diminish progressively along the chain. For example, the transition frequency for the 1st through 4\textsuperscript{th} infection AEs ranged 300-420; but beyond the 4\textsuperscript{th} infection event, the frequency reduced to the range of 50-300. The transition probabilities from bleeding to infection were 39\% to 45\%, while from infection to bleeding were less than 18\%. There were also 285 patients (8\%) who received a heart transplant after experiencing an infection AE (represented by the dark green circle in the 2\textsuperscript{nd} column in Fig. 12). There were also transitions from cardiac arrhythmia and respiratory failures as the 1st AEs to infection as the 2\textsuperscript{nd} AEs.

		\item \textbf{\textit{GRP4: Trajectories to transplant}}

		The Markov chains for GRP4  (Fig. 13) represents AE trajectories of 3,302 patients who ended in receiving a heart transplant. The majority of patients who revived heart transplants was precede by various types of AE, most commonly bleeding, infection, and cardiac arrhythmia. Only 970 patients (29\%) in this group received a heart transplant as the initial, isolated event (the dark green circle in the first column of Fig. 13). AE trajectories to heart transplants from some specific types of AE like bleeding, infection, or cardiac arrhythmia were more likely than other types of AE like neurological dysfunction AE or device malfunction AE. For instance, 520 patients who experienced only one AE and then received a heart transplant were preceded by mostly bleeding or cardiac arrhythmia (cumulative 388 patients), and minimally by neurological dysfunction or device malfunction (cumulative 132 patients).

		\begin{figure}[h]
			\centering
			\includegraphics[width=0.4\textwidth,height=0.18\textheight]{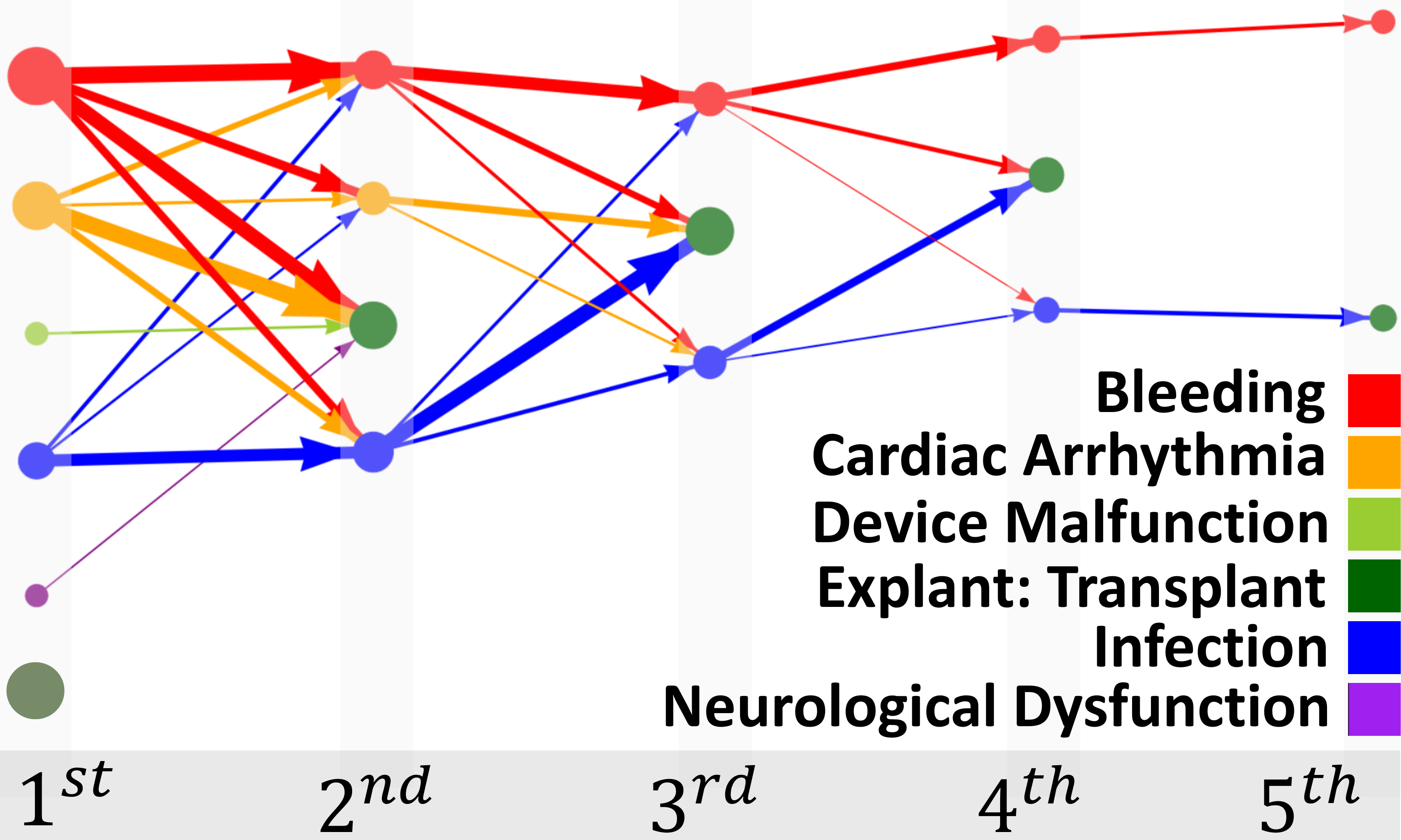}
			\caption{MM of GRP4: Trajectories to transplant (n= 3,302 patients)}
		\end{figure}

		\item \textbf{\textit{GRP5: Cardiac arrhythmia}}

		Fig. 14 represents a chain of recurrent cardiac arrhythmia AEs (orange circles) in GRP5 (1,275 patients), with a limited amount of branching involving an intermediate infection AE (blue circles). The frequency of transitions depicted by the thickness of the arrows is seen to diminish progressively along the chain. As an example, 919 patients in GRP5 (72\%) experienced cardiac arrhythmia as 1\textsuperscript{st} AE and only 242 of them experienced cardiac arrhythmia as the 2\textsuperscript{nd} AE too; beyond the 2\textsuperscript{nd} AE the frequency gradually decreased to 71 for cardiac arrhythmia as the 6\textsuperscript{th} AE. The probability of transitions for recurrent cardiac arrhythmia AEs were between 34\% to 49\% over the time. There were also transitions between cardiac arrhythmia and infection with probabilities from 16\% to 36\%.

		\begin{figure}[h]
			\centering
			\includegraphics[width=0.4\textwidth,height=0.09\textheight]{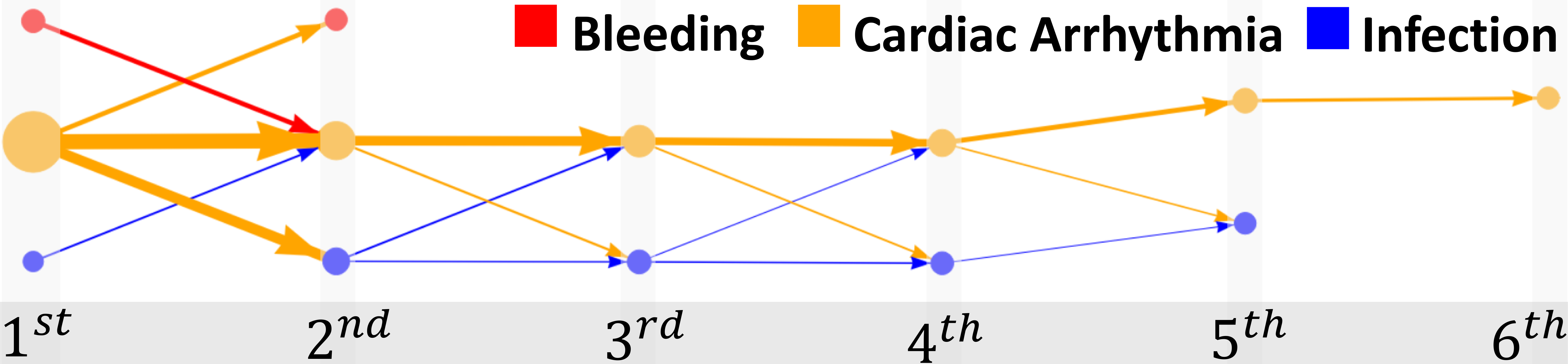}
			\caption{MM of GRP5: Cardiac arrhythmia (n= 1,275 patients)}
		\end{figure}

		\item \textbf{\textit{GRP6: Trajectory of neurological dysfunction \& death }}

		The Markov chain of GRP6 (Fig. 15) shows the trajectory of 1,616 patients who died (yellow circles) after suffering from neurological dysfunction AEs (purple circles). The rate of death for patients who suffered from neurological dysfunction AEs as the 1\textsuperscript{st}  AE trough 3\textsuperscript{th} AE ranged from 27\% to 39\%. The majority of patients for which neurological dysfunction AE was recorded was preceded by other types of AE, most commonly bleeding and infection. Only a small number of patients in GRP6 (n=229, approximately 14\%) died with no reported AEs (the yellow circle in the 1\textsuperscript{st} column of Fig. 15). There were also a small number of patients who died after one or recurrent infection AEs with no recorded neurological dysfunction AE (thin blue arrows from blue circles to yellow circles).

		\begin{figure}[h]
			\centering
			\includegraphics[width=0.4\textwidth,height=0.15\textheight]{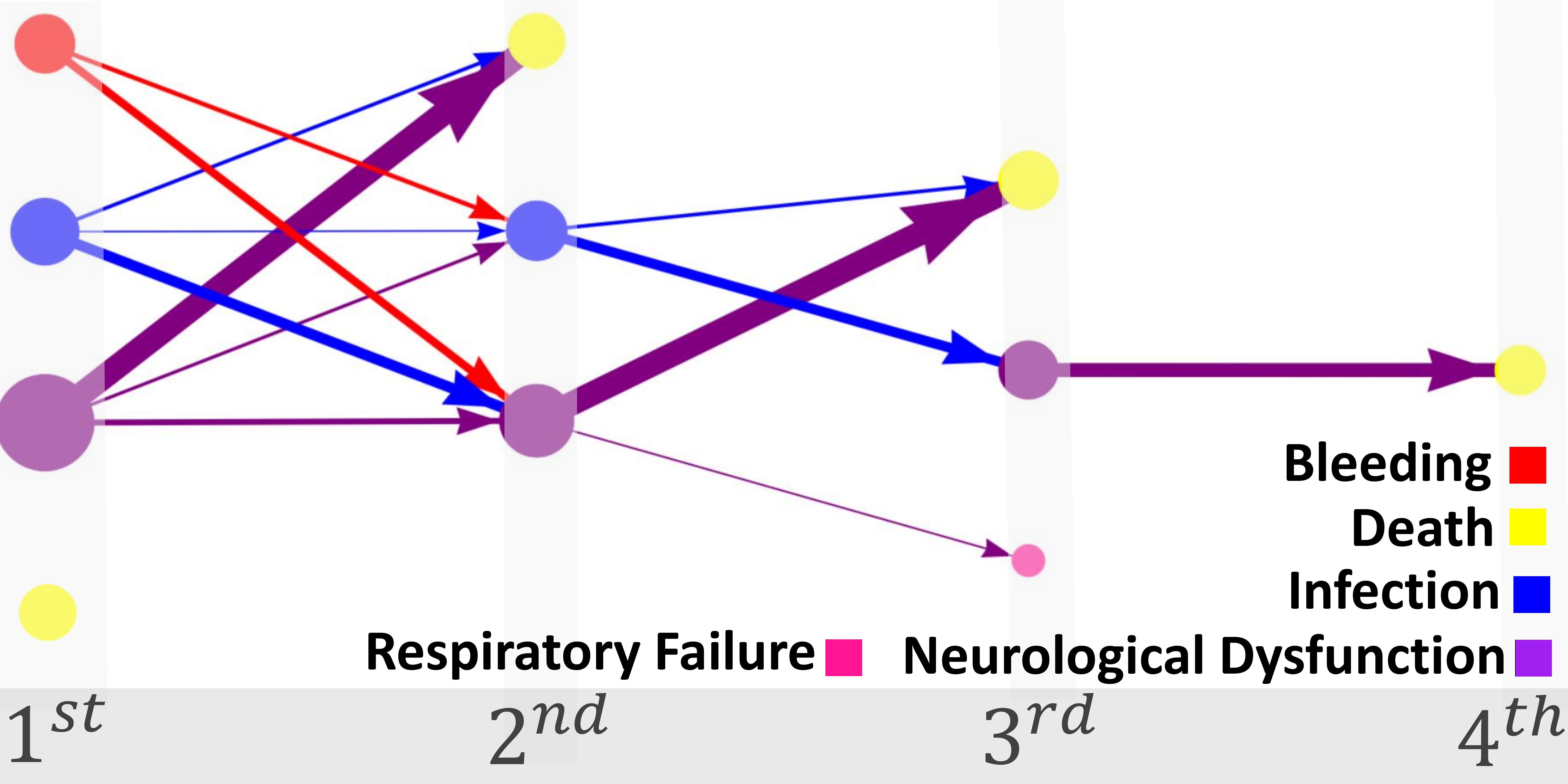}
			\caption{MM of GRP6: Trajectory of neurological dysfunction \& death (n= 1,616 patients)}
		\end{figure}

		\item \textbf{\textit{GRP7: Trajectory of respiratory failure \& renal dysfunction \& death}}
		
		The Markov chain of GRP7 (Fig. 16) illustrates two different AE trajectories to death. One main trajectory represents 929 patients who died after suffering from respiratory failure (934 recorded respiratory failure AE) and/or renal dysfunction AEs (712 recorded renal dysfunction AEs) with transition probabilities between 22\% to 44\%. It also revealed that patients with reported respiratory failure and renal dysfunction  were preceded by other types of AE, most commonly by infection AE and bleeding AE. The transition probabilities from renal dysfunction or infection to respiratory failure ranged from 32\% to 36\%. Another trajectory of this group presents 179 patients who died after suffering from one or recurrent bleeding AEs with transition probabilities from 19\% to 24\%.

		\begin{figure}[h]
			\centering
			\includegraphics[width=0.4\textwidth,height=0.15\textheight]{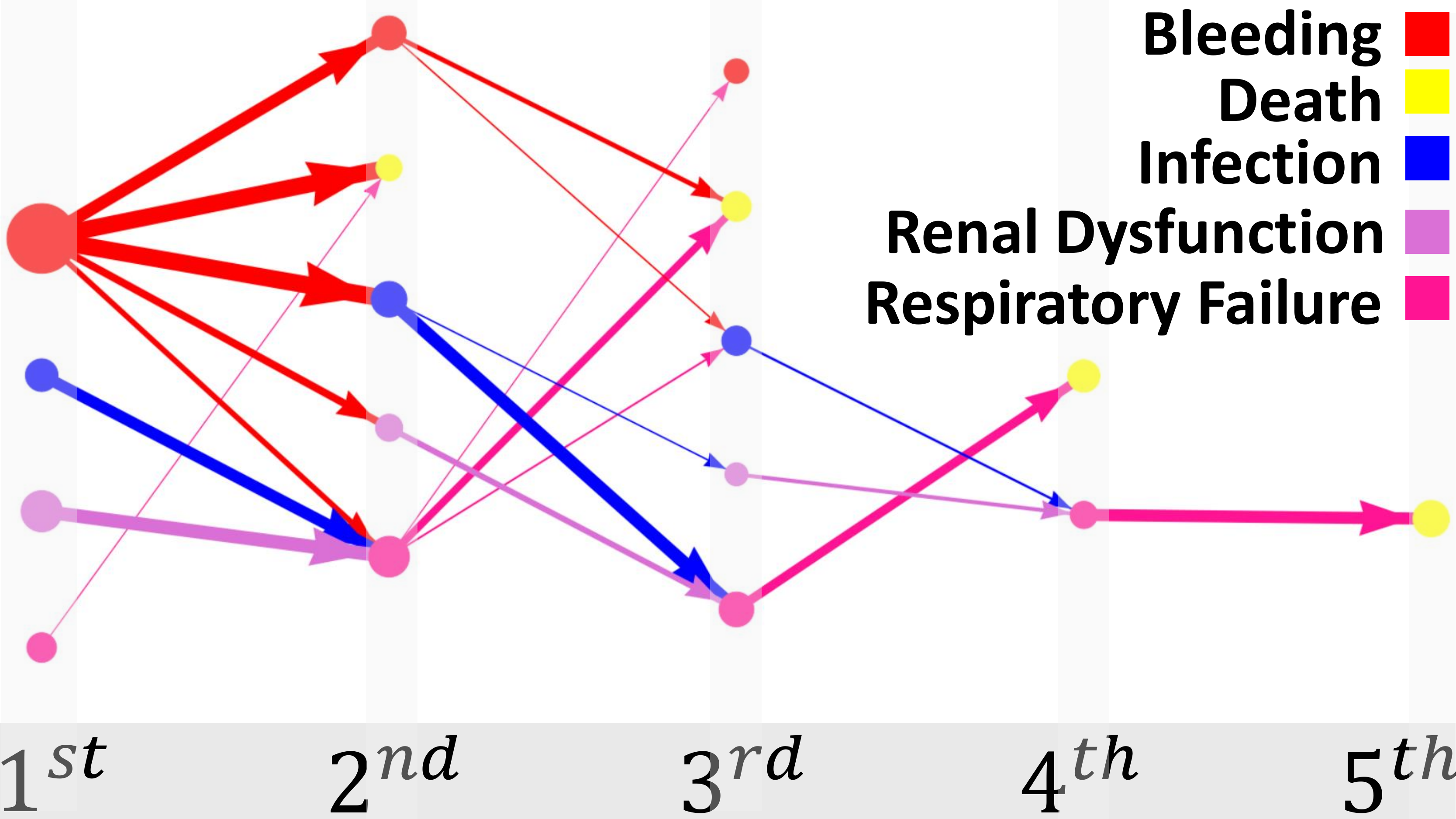}
			\caption{MM of GRP7: Trajectory of respiratory failure \& renal dysfunction \& death (n= 1,108 patients)}
		\end{figure}

	\end{itemize}

	\begin{table*}[h]\centering
		\caption{Summary of Clinical Insights of the Groups Resulted from Hierarchical Clustering.}
		\renewcommand{\arraystretch}{1.5}
		\footnotesize
		\begin{tabular}{c}
			\hline
			\rowcolor{gray!10}  
			\textcolor{red}{\textbf{GRP1:Recurrent bleeding}}\\ \hline
			\parbox[t][0.65cm][t]{18cm} {Patients who experienced recurrent bleeding AEs after LVAD. Recurrent occurrences of various types of bleeding such as  gastrointestinal bleeding is commonly reported in clinical studies of patients who receive an LVAD \cite{bunte2013major,sami2014gastrointestinal,draper2014gi}}\\\hline
			
			\hline
			\rowcolor{gray!10}
			\textcolor{green}{\textbf{GRP2:Trajectory of device malfunction \& explant}}  \\ \hline
			\parbox[t][1.9cm][t]{18cm} {Patients who had device malfunction and commonly had their LVADs removed (explanted). This trajectory was found to be preceded by the two types of AEs including infection and bleeding. Two others, less common, trajectories within GRP2 were 1. patients who had an explant without device malfunction and 2. patients who had a device malfunction without ending in explant. In clinical literature, device malfunction is defined as failure of one or more parts of LVAD that cause the inability to maintain adequate circulatory support. Device malfunction might be deadly or could be solved by replacing the device (explant) \cite{dembitsky2004left,rizzieri2008ethical}. It is important to note that within INTERMACS patients there may be some who had serious pump malfunction such as internal thrombosis, but the patient was not considered a candidate for pump exchange and the LVAD may have been simply turned off.}
			
			\\\hline

			\rowcolor{gray!10}
			\textcolor{blue}{\textbf{GRP3:Infection}}    
			\\\hline
			\parbox[t][0.65cm][t]{18cm} { Patients  who suffered mostly from infection AEs. The most recent INTERMACS annual report indicated infection as the most frequent AE after bleeding during the first three months and the most common AE thereafter \cite{kirklin2017eighth}.}
			
			\\\hline

			\rowcolor{gray!10}
			\textcolor{OliveGreen}{\textbf{GRP4:Trajectories to heart transplant}}    
			\\\hline
			
			\parbox[t][1.56cm][t]{18cm} {Patients who received a heart transplant and the pump was explanted as part of the procedure. AE trajectories to heart transplant were mostly consisted of bleeding AE, infection AE, and cardiac arrhythmia AE. In practical terms these AEs resulted in an upgraded listing for cardiac transplant resulting in a higher likelihood of achieving a heart transplant. It is not uncommon for LVAD complications to drive more urgent listing status of a candidate and this analysis conforms that strategy. The most recent INTERMACS annual reports indicated slightly more than 30\% of heart transplant candidates who received continuous-flow LVAD  received a  heart transplant \cite{kirklin2015seventh,kirklin2017eighth}.}
			
			\\\hline
			
			\hline
			\rowcolor{gray!10}
			\textcolor{orange}{\textbf{GRP5:Cardiac arrhythmia}}  
			
			\\\hline
			\parbox[t][0.65cm][t]{18cm} {Patients who experienced one or recurrent cardiac arrhythmia AEs after LVAD implant that accompanied mostly with infection AEs and bleeding AEs. Patients with ventricular arrhythmias tend to have recurrent episodes of these rhythm disturbances one they begin to manifest them.}

			\\\hline
			
			\rowcolor{gray!10}
			\textcolor{Plum}{\textbf{GRP6:Neurological dysfunction \& death}}  
			
			\\\hline
			\parbox[t][0.65cm][t]{18cm} {GRP6 trajectory is supported by clinical literature highlighting the high mortality in LVAD patients with neurological events especially with hemorrhagic strokes \cite{al2016neuroendovascular,harvey2015stroke,willey2016outcomes,kato2012pre}}
			
			\\\hline
			
			\rowcolor{gray!10}
			\textcolor{Magenta}{\textbf{GRP7:Trajectory of respiratory failure \& renal dysfunction \& death}} 
			
			\\\hline
			\parbox[t][0.5cm][t]{18cm} {GRP7 trajectory is predominated by both  respiratory failure AEs and renal dysfunction AEs that ended in death although some of the trajectories also were accompanied by a bleeding AE and an infection AE. The eighth INTERMACS annual report named renal dysfunction and chronic pulmonary disease among the non-cardiac system commodities that impact the LVAD survival rate  \cite{kirklin2017eighth}.}

		\end{tabular}
		
	\end{table*}
	
	\section{Discussion}
	
	This is the first study to explore patterns of sequential AEs and their related final outcomes in patients with advanced heart failure who received an LVAD implant. The results of our analysis revealed  previously-unknown or under-appreciated patterns of adverse events in LVAD patients, and their relationships with final outcomes, as presented in Table. \rom{3}. These trajectories can be used as the basis for personalized medical management of the LVAD patient: by identifying developing patterns of AE’s and therefore intervening with evasive treatment.
	

	Patients were not equally distributed among the groups that resulted from clustering. As an example, \textit{GRP1 (Recurrent bleeding)} include small group of patients (less than 1000) who experienced greatest  numbers of AEs (with a median of 9 AEs).  While, each of \textit{GRP3 (Infection) } and \textit{GRP4 (Trajectories to transplant)} included more than 3000 patients with the median number of 3 AEs. The imbalanced number of patients in the groups implies high incidence of specific AE patterns among the patients with LVAD that seek more clinical attention for future enhancement of LVAD outcome.

	The time of occurrence following implant is very important to the analysis of AEs. It is highly valuable for physicians to know how many days or months after LVAD implant AEs are likely to occur or how quickly a series of AEs may occur for a patient. The time analysis of AE sequential patterns (Table. \rom{2}) indicates various timing characteristics among the groups. For instance, AEs in the GRP7  with respiratory and renal failure occurred in the immediate  months after LVAD implant with median time of 1 month after LVAD implant reflecting typical post-operative timing of these events. This is in contrast to AEs in the bleeding group (GRP1) with median time of 9 months  after LVAD implant, more typical of onset of gastrointestinal bleeding. This analysis also revealed differences in time span over which AEs occurred between groups. For example, AEs for patients with neurological events  (GRP6) occurred over a short period of time (median time span of 1 month) contrasted with GRP5 patients whose AEs spanned a median of 4 months. These different timing characteristics of AE sequential patterns may have implications in guiding post-LVAD medical interventions or preventative measures.\\

	\noindent
	\textbf{\textit{Challenges \& Limitations \& Future work}}

	The first challenge encountered in this study was related to data preprocessing including choosing a right population of patients and AEs. This part of study was an iterative process, involving feedback from clinicians about medical interpretation of the results. As an example, it was decided to exclude AEs after the explant of the first LVAD as patterns of AEs and clinical decisions and therapies for patients with more than one device are unique and need to be studied separately. Also, the choice of criteria to evaluate clustering results and decide about the number of groups was another challenge. Zhang et al. \cite{zhang2015paving} used Silhouette values to determine cluster number and support values for internal evaluation of clustering. However, the number of patients in this study (13,192 patients) was more than 10 times greater compared with the number of patients in \cite{zhang2015paving} (1,576 patients). Thus, considering only support values was inadequate. It is obvious that high numbers of groups  will result in increasing similarity of patients within each group; however, the number of groups should be limited to preserve the clinical utility of the results. This was the motivation to the step-wise clustering evaluation that was implemented that considers both within-group similarity and between-groups dissimilarity criteria.
	
	Markov models in this study considered the order of occurrences for AE transitions in patient's sequences. For instance, the transition from bleeding as the first AE to infection as the second AE was considered different from transition from bleeding as the fifth AE to infection as the sixth AE. Thus, the probabilities of transitions were evaluated by considering where in the patient's sequence transitions occurred. This temporal constraint is very useful since AEs have a different effect on LVAD final outcome when they occurred in immediate succession as compared to a sequence with a different intermediate AE.
	
	One main limitation of this study is related to the voluntary collection and reporting of INTERMACS data. As an example, some AEs like infection is a longitudinal AE that might last for a while, but INTERMACS only records the occurrence of AE without recording its duration. Another issue is that there is no information regarding the order of concurrent AE (events occurred at the same day). The problem was exacerbated by the fact that this registry is comprised of contributions of over 100 centers, and hence involves differences in interpretation of definitions, omissions, and data entry errors. As an example, the ongoing change in the definition of right heart failure (RHF) causes inconsistency between studies about analysis of RHF, and therefore, reduces the confidence in results. Consequently, it would be helpful to pull in expertise from field to learn more about INTERMACS definitions and workflow of data collection, to avoid bias in the future studies.
	
	The time gaps between sequential AEs were not considered in clustering and Markov modeling to minimize the diversity of sequences. One solution that was evaluated was to segment post-LVAD time, based on the critical points based on the AEs distribution and physicians' suggestions. This reduced the maximum number of time points in the sequences from 36 to 7. However, it was not adequate to prevent enormously increasing diversity. Further consolidating the timeline into short-term and long-term could be another solution that requires some iterative process to find an optimum. Adding time gaps will also help concurrent AEs issue by considering basket of AEs with 0 time gaps as one element in the sequence\cite{ayres2002sequential}.  
	

	The existence of sequential post-LVAD AEs raises the question of what causes patients to experience different AEs patterns.  Further research of post-LVAD AEs must seek to include pre and post LVAD risk factors to improve the prognostic value of this analysis. An accurate prediction of the next AE (AE\textsubscript{n+1}) by considering previous AEs (AE\textsubscript{1} to AE\textsubscript{n}) would be another important contribution, and a valuable tool for physicians to optimize treatment to minimize the risk of future AEs.
	
	\section{Conclusion}
	
	This study, to the best knowledge of the authors, was the first exploratory to discover sequential chains of AEs following LVAD implant. Mining of the AE sequences of 13,192 patients with advanced heart failure derived from the INTERMACS registry revealed the existence of seven groups of sequential chains of AEs, each characterized by a dominant AE or multiple AEs and occurring in a unique order. The discovered chains of AEs disclose potential interdependence between AEs and provide clinicians a valuable insight into the patient oriented post-LVAD AEs evidence. It is hoped that this analysis may support post-LVAD follow-up by alerting medical providers of the likelihood of impending AEs - based on a combination of independent factors, and patterns of prior AEs.


	
	

	\section{Acknowledgments}
	This work was supported by an R01 grant (R01HL086918) from the National Institutes of Health/National Heart, Lung, and Blood Institute. We would also like to thank the Data Access, Analysis, and Publications Committee of INTERMACS for allowing us to use their registry for the study.

	\bibliographystyle{IEEEtran}
	\bibliography{libraryabv}	
\end{document}